\documentclass[journal,draftclsnofoot,onecolumn]{IEEEtran}

\usepackage{amssymb}
\usepackage{bm}
\usepackage{cancel}
\usepackage[table]{xcolor}
\usepackage{multicol}
\usepackage{cancel}
\usepackage{caption,subfig}
\ifCLASSOPTIONcompsoc
  \usepackage[nocompress]{cite}
\else
\usepackage{cite}
\fi

\ifCLASSINFOpdf
\else
\fi

\usepackage[cmex10]{amsmath}
\usepackage{amsthm}
\usepackage{paralist}
\usepackage{multirow}
\usepackage{booktabs}
\usepackage{threeparttable}
\usepackage{algorithm}
\usepackage[noend]{algorithmic}
\newtheorem{lemma}{Lemma}
\newtheorem{theorem}{Theorem}

\theoremstyle{definition}
\newtheorem{definition}{Definition}
\newtheorem{example}{Example}

\newcommand{\bx}{\mathbf{x}}
\newcommand{\by}{\mathbf{y}}
\newcommand{\bY}{\mathbf{Y}}
\newcommand{\rbracket}{]}
\newcommand{\range}[2]{#1 : #2}
\newcommand{\mf}[1]{\mathbf{#1}}
\newcommand{\iters}{s}

\begin{document}
\title{Decoding and Repair Schemes for Shift-XOR Regenerating Codes}

\author{Ximing~Fu, Shenghao~Yang and Zhiqing~Xiao % <-this % stops a space
\IEEEcompsocitemizethanks{
\IEEEcompsocthanksitem This paper was presented in part at TrustCom 2014 and ISIT 2015.
\IEEEcompsocthanksitem
Ximing Fu is with the School of Science and Engineering, The Chinese University of Hong Kong, Shenzhen, Shenzhen, Guangdong 518172 and University of Science and Technology of China, Hefei, Anhui 230052, China (e-mail: fuxm07@foxmail.com).
\IEEEcompsocthanksitem S. Yang is with the School of Science and Engineering and the Shenzhen Key Laboratory of IoT Intelligent Systems and Wireless Network Technology, The Chinese University of Hong Kong, Shenzhen, Shenzhen Guangdong 518172, China, and also with the Shenzhen Research Institute of Big Data, Shenzhen, Guangdong 518172, China (e-mail: shyang@cuhk.edu.cn).
\IEEEcompsocthanksitem
Zhiqing Xiao was with the Tsinghua University when participating in the research of this paper (e-mail: xzq.xiaozhiqing@gmail.com).}% <-this % stops an unwanted space
}
%\thanks{This work was supported by\ldots }}%Manuscript received April 19, 2005; revised September 17, 2014.}}

\IEEEtitleabstractindextext{%
  \begin{abstract}
    Decoding and repair schemes are proposed for shift-exclusive-or
    (shift-XOR) product-matrix (PM) regenerating codes, which
    outperform the existing schemes in terms of both communication and
    computation costs. In particular, for the shift-XOR minimum
    bandwidth regenerating (MBR) codes, our decoding and repair
    schemes have the optimal transmission bandwidth and can be
    implemented in-place without extra storage space for intermediate
    XOR results. Technically, our schemes involve an in-place
    algorithm for solving a system of shift-XOR equations, called
    \emph{shift-XOR elimination}, which does not have the bandwidth
    overhead generated by shift operations as in the previous
    zigzag algorithm and has lower computation complexities compared
    with the zigzag algorithm. The decoding
    and repair of shift-XOR MBR/MSR codes are decomposed into a
    sequence of systems of shift-XOR equations, and hence can be
    solved by a sequence of calls to the shift-XOR elimination. As the
    decompositions of the decoding and repair depend only on the PM
    construction, but not the specific shift and XOR operations, our
    decoding and repair schemes can be extended to other MBR/MSR codes
    using the PM construction.  Due to its fundamental role, the
    shift-XOR elimination is of independent interest. 
\end{abstract}

}
\maketitle
\IEEEdisplaynontitleabstractindextext

\IEEEpeerreviewmaketitle

\section{Introduction}
\label{sec::intro}
Distributed storage systems with potential node failures usually use
redundancy to ensure the reliability of the stored data. Compared with repetition, erasure coding
is a more efficient approach to introduce redundancy. Using an
$\left[n,k\right]$ maximum-distance separable (MDS) erasure code, a
data file of $kL$ bits is divided into $k$ sequences, each of $L$ bits.  The $k$ sequences are encoded into $n$ coded sequences and
stored in $n$ nodes, each storing one coded sequence. A decoder
can decode the data file from any $k$ out of the $n$ coded sequences.
An $\left[n,k\right]$ MDS erasure code can tolerate at most $n-k$
node failures. Reed-Solomon codes~\cite{reed1960polynomial} are
widely used MDS erasure codes, where the encoding and decoding operations are over finite fields, and have high computation costs.
% Towards reducing the computation costs, sparse parity-check was explored to reduce the number of multiplications~\cite{BlaumR99}.

Towards low complexity codes, exclusive-or (XOR) and cyclic-shift
operations have been employed to replace finite-field operations.  One
family of such MDS storage codes includes EVENODD codes
\cite{blaum1995evenodd} and RDP codes~\cite{corbett2004row} for
tolerating double node failures, and their
extensions~\cite{blaum1996mds,blaum2001evenodd,HuangX08} for
tolerating triple or more node failures. These codes share a common
cyclic-shift Vandermonde generator matrix and the decoding complexity
of these codes has been improved by LU factorization of Vandermonde
matrix~\cite{HouHSL18}.  Another family of MDS storage codes using XOR and
cyclic-shift is based on Cauchy generator matrices, including Cauchy
Reed-Solomon \cite{bloemer1995xor,plank2006optimizing} and Rabin-like
codes~\cite{feng2005new2}, where the decoding method is improved
in~\cite{HouH18Tcom}.
% complexity is as high as $O\left((n-k)^4L^3\right)$, which was improved to $O((n-k)^4L)$

% Using XOR operation, 
% XOR and cyclic-shift operations were applied to construct codes to tolerate double \cite{Plank08} and triple~\cite{LinWSLL10} node failures.

% Blaum-Roth codes~\cite{BlaumR93}

% 
% with low complexity,
% using  which can be more efficiently implemented
% than finite-field operations.

In this paper, we focus on a class of storage codes based on
(non-cyclic) shift and XOR operations, called shift-XOR
codes. Sung and Gong~\cite{sung2013zigzag} presented a class of
storage codes for any valid pair $\left[n,k\right]$ using shift and
XOR operations, where the generator matrix satisfies the \textit{increasing
  difference property}. Using an $[n,k]$ shift-XOR storage code,
the $kL$ bits of the data
file can be decoded from any $k$ out of the $n$ coded sequences using 
the zigzag decoding algorithm~\cite{sung2013zigzag}.
Due to shift operations, the coded
  sequences are usually longer than $L$ bits so that the shift-XOR
  storage codes are not strictly MDS, where the extra bits are called
  the \emph{storage overhead}. An $[n,k]$ shift-XOR storage code is asymptotic MDS when $L$ is large. Moreover, the total number of bits
retrieved by the zigzag decoder is more than $kL$ bits, where the extra
bits are called the \emph{(decoding) bandwidth overhead}.

Shift-XOR codes have attracted more research interests recently due to
the potential low encoding/decoding computation costs. The
shift-XOR storage codes with zigzag decoding have lower encoding and
decoding complexities than Cauchy Reed-Solomon codes in a wide range
of coding parameters \cite{gong17zigzag}. A fountain code based on
shift and XOR outperforms Raptor code in terms of the
transmission overhead \cite{nozaki2014fountain}. Efficient repair
schemes for shift-XOR storage codes have been studied in
\cite{dai17new}.  Moreover, shift and XOR operations can also be used
to construct network codes~\cite{sung2014combination} and
\emph{regenerating codes}~\cite{hou2013basic}.

For distributed storage system, it is also worthwhile to consider the
\emph{repair of failed nodes}. Dimakis \emph{et al.} formulated
regenerating codes to address this issue~\cite{dimakis2007network}.
In an $\left[n,k,d\right]$ regenerating code, a data file of $BL$ bits
is divided into $B$ sequences, each containing $L$ bits.  The
sequences are encoded into $n\alpha$ sequences each of $L$ bits and
distributed to $n$ storage nodes, each storing $\alpha$ sequences of $L$ bits.
The data file can be decoded from any $k$ storage nodes, and a
failed node can be repaired from any other $d$ surviving nodes.  There
are two kinds of repair~\cite{dimakis2007network}: \textit{exact
  repair} and \textit{functional repair}.  In exact repair, the
sequences stored in the failed node can be exactly reconstructed in
the new node.  In functional repair, the sequences reconstructed in
the new node may be different from those in the failed node as long as
the new node and the other nodes form an $[n,k,d]$ regenerating code.

The tradeoff between the storage in a node and the repair bandwidth is
characterized in~\cite{dimakis2007network}.  Two classes of codes that
achieve the optimal storage-repair-bandwidth tradeoff are of particular
interests, i.e., the minimum bandwidth regenerating (MBR) codes and
minimum storage regenerating (MSR) codes.  Rashmi, Shah and Kumar~\cite{rashmi2011optimal} proposed product-matrix (PM)
constructions of MBR codes for all valid tuples $\left[n,k,d \right]$ and
MSR codes for $d \ge 2k-2$ with exact repair algorithms. 
In~\cite{rashmi2011optimal}, PM MSR codes with $d>2k-2$ are constructed based on the construction of the $d=2k-2$ case. Two unified constructions of MSR codes for $d\geq 2k-2$ were proposed in~\cite{LinCHA15,KuriharaK13}. Based on special parameterized codes such as determinant codes with $k=d$~\cite{ElyasiM19}, Cascade codes were proposed to achieve the MSR tradeoff point with arbitrary feasible combinations of $n,k,d$~\cite{Elyasi19Cascade}. Some MDS codes with sub-packetization were proposed to construct MSR codes~\cite{YeB17,YeB17a,LiTT18}.

The PM MBR/MSR codes require matrix operations over finite fields for
encoding, decoding and repair, which have the high computation cost as
Reed-Solomon codes.  To achieve lower complexity, Hou \emph{et
  al.}~\cite{hou2013basic} proposed the regenerating codes using shift
and XOR operations based on the PM construction. The shift-XOR MBR/MSR
codes have the storage overhead the decoding/repair bandwidth
overhead. For example, for the shift-XOR MBR codes, extra
$\frac{1}{2}k(k-1)$ sequences are retrieved for decoding, where the
sequence length is at least $L$. Hou \emph{et al.}~\cite{HouSCL16}
proposed another class of regenerating codes using cyclic-shift and
XOR based on the PM construction, called \emph{BASIC codes}, and
demonstrated the lower computation costs than the finite-field PM
regenerating codes.  The BASIC MBR codes have the similar decoding
bandwidth issue as the shift-XOR regenerating codes.  % The
% finite-field and cyclic-shift PM regenerating codes have no
% storage overhead.

% The decoding and repair
%   algorithms in~\cite{hou2013basic} use matrix operations, that have high computation costs and have the decoding/repair bandwidth overhead: i.e., more bits are retrieved than the number of bits to decode/repair.

% Paying the storage overhead, could shift-XOR regenerating codes gain in
% bandwidth overhead and computation complexities?
In~\cite{hou2013basic}, a general sufficient condition is provided
such that a system of shift-XOR equations is uniquely solvable, which
induces an adjoint matrix based approach to solve a system of
shift-XOR equations. However, the adjoint matrix based approach has a
high computation cost so that the decoding and repair computation
costs of the shift-XOR regenerating codes in~\cite{hou2013basic} are
higher than those of the BASIC codes in~\cite{HouSCL16} (see the
comparison in Table~\ref{tab::compare}). In this paper, we will show that
it is possible to reduce the decoding/repair complexities of the
shift-XOR regenerating codes to be better than or similar to those of
BASIC codes.

\subsection{Our Contributions}

In this paper, we first study solving a system of shift-XOR equations,
where the generator matrix satisfies a refined version of the
increasing difference property (see Section~\ref{sec::pre}). The
\emph{refined increasing difference (RID)} property relaxes the
original one in~\cite{sung2013zigzag} so that the storage overheads
can be smaller. The RID property is satisfied by, for example,
Vandermonde matrices. We propose an algorithm, called \emph{shift-XOR
  elimination}, for solving such a system of shift-XOR equations (see
Section~\ref{sec::mds}).  Our algorithm has the following properties:
\begin{itemize}
\item Bandwidth overhead free: using the shift-XOR elimination to decode a shift-XOR storage code, only subsequences stored in a storage node are needed, and the number of bits retrieved from the storage nodes is equal to the number of bits to decode. In other words, the bandwidth costs of the shift-XOR elimination is optimal.
\item Lower computational space and time complexities: the shift-XOR
  elimination has a smaller number of XOR operations and smaller
  auxiliary space compared with the zigzag algorithm. The number of
  XOR operations used by the shift-XOR elimination is the same as the
  number of XOR operations used to generate the input subsequences
  from the message sequences. The shift-XOR elimination only needs a
  constant number of auxiliary variables, and can be implemented
  in-place: the results of the intermediate XOR operations and the
  output bits are all stored at the same storage space of the input
  binary sequences.  In other words, the algorithm uses an auxiliary
  space of $O(1)$ integers per $kL$ bits to solve, and hence has an
  asymptotically optimal space cost as $kL$ is large.
\end{itemize}
In Table~\ref{tab:dec}, the shift-XOR elimination and the zigzag
algorithm are compared for decoding a shift-XOR storage code.  The
shift-XOR elimination can be used in other shift-XOR codes, e.g.,~\cite{nozaki2014fountain,dai17new}. In this
paper, we focus on its application in regenerating codes.

\begin{table}[b]
  \centering
  \caption{Comparison of the algorithms for decoding a shift-XOR storage code in terms of bandwidth overhead, decoding auxiliary space, and decoding complexities. Here $L$ is the message sequence length, and $k$ is the number of message sequences. A Vandermonde generator matrix is assumed.}
  \label{tab:dec}
  \begin{tabular}{ccccc}
    \toprule
    \multirow{2}{*}{Decoding Algorithm} & \multirow{2}{*}{Bandwidth Overhead} & Decoding & Decoding & Decoding \\
      &   & Auxiliary Space &  XOR Operation & Integer Operation \\
    \midrule
    {shift-XOR elimination (Section~\ref{sec::mds})} & $0$ & $O(1)$ integers  & $<{k}(k-1)L$ & $O(k^2L)$ \\
     zigzag decoding \cite{sung2013zigzag} & $i\left( k-1 \right)$ bits for node $i$ & \parbox[t]{20mm}{\centering $O\left( kL \right)$ integers\\and $kL$ bits} & $k(k+1)L$ & $O({k}^{2}L)$ \\
    \bottomrule
  \end{tabular}
\end{table}

% we study the decoding and repair schemes for shift-XOR
%  regenerating codes. Compared with the best existing shift-XOR decoding/repair
%  schemes~\cite{hou2013basic,HouSCL16}, ours are more efficient in
% terms of communication bandwidth, auxiliary space and time complexity. The contributions of this paper are summarized as follows.

In Section~\ref{sec::mbr} and~\ref{sec::msr}, we study decoding and repair of the shift-XOR PM regenerating codes proposed in~\cite{hou2013basic}. Our decoding/repair schemes transform the decoding/repair problem to a sequence of subproblems of shift-XOR equations, which can be solved using the shift-XOR elimination. Benefit from the advantages of the shift-XOR elimination, our schemes in general have lower computation and bandwidth costs than those in~\cite{hou2013basic} (see Table~\ref{tab::compare}).

In particular, for the shift-XOR MBR codes, our decoding and repair schemes retrieve the same number of bits as the sequences to decode or repair. The decoding and repair schemes can be implemented in-place with only $O(1)$ integer auxiliary variables, but without any auxiliary variables to store the intermediate XOR results. Moreover, for decoding, the number of XOR operations is the same as the number of XOR operations for generating the input subsequences from the message sequences.
For the shift-XOR MSR codes, our decoding and repair schemes have the same bandwidth cost, smaller auxiliary space for intermediate XOR results, and lower order of the number of XOR operations than those of~\cite{hou2013basic}.

With our decoding and repair schemes, shift-XOR regenerating codes can have better or similar performance compared with the BASIC codes~\cite{HouSCL16}, which use cyclic-shift and XOR. In particular, for MBR decoding and repair and MSR repair, our schemes have lower auxiliary spaces; for MBR decoding, our scheme has a smaller number of XOR operations. When $L$ is sufficiently larger than $nd$, for MBR repair and MSR decoding and repair, our schemes have a similar or smaller number of XOR operations. See the comparison in Table~\ref{tab::compare}.

\begin{table}[tb]
  \caption{Comparison among Decoding and Repair Schemes of (cyclic-)shift-XOR MBR/MSR Codes. Here $B$ is the number of message sequences, $L$ is the message sequence length, $k$ is the number of nodes for decoding, $n$ is the number of storage nodes, $d$ is the number of helper nodes for repair, and $i$ is the node to repair. For MSR codes, $d=2k-2$. A Vandermonde generator matrix is assumed. % For repairing the sequences at node $i$, the $d$ helper nodes are indexed with $i_1,i_2,\ldots,i_d$.
%The repair of node $i$ is considered.
}
  \label{tab::compare}
  \addtolength{\tabcolsep}{-4pt}
  \centering
  \begin{tabular}{ccccc}
    \toprule
    \multicolumn{2}{c}{Algorithm} & Bandwidth  & \parbox[t]{20mm}{\centering Auxiliary Space \\for XOR results} & XOR Operation \\
    \midrule
    \multirow{3}{*}{MBR Decoding} & {this paper (Section~\ref{sec::mbr_decode})} & {$BL$} & $0$ &  $O(dk^2L)$ \\
& \cite{hou2013basic} & {$>BL+\frac{k(k-1)}{2}L$} & {$>BL+\frac{k(k-1)}{2}L$} & $> {d}^{2}{k}^{3}nL $ \\
& \cite{HouSCL16} & {$>BL+\frac{k(k-1)}{2}L$} & {$>BL+\frac{k(k-1)}{2}L$} & { $O(dk^{3}L)$}\\
    \midrule

    \multirow{3}{*}{MBR Repair} & {this paper (Section~\ref{sec::mbr_repair})} & {$d(L+(i-1)(d-1))$} & $0$ & $2d(d-1)L+O(nd^3)$  \\
& \cite{hou2013basic} & {$d(L+(i-1)(d-1))+ O(nd^2)$}  & {$>dL$} & $> d^{4}nL $ \\
& \cite{HouSCL16} & {$dL$} & {$dL$} &  {$\left(\frac{15}{4}d-\frac{3}{4}\right)dL$} \\
 \midrule
    \multirow{3}{*}{MSR Decoding} & {this paper (Section~\ref{sec::msr_decode})} & {$k(k-1)L+O(nk^2d)$} & {$(k-1)(k-2)L+O(nk^2d)$} & $O(k^3L+nk^3d)$ \\
& \cite{hou2013basic} & {$k(k-1)L+O(nk^2d)$} & {$k(k-1)L+O(nk^2d)$} & $> {d{k}^{4}n}L $ \\
& \cite{HouSCL16} & {$k(k-1)L$} & {$k(k-1)L$} & {$O(k^{4}L)$}\\
\midrule
    \multirow{3}{*}{MSR Repair} & {this paper (Section~\ref{sec::msr_repair})} & {$dL+O(nd^2)$} & {$\frac{1}{2}dL+O(nd^2)$} & $\frac{3}{2}\left(d-1\right)dL+O(nd^3)$ \\ %$O(d^2L+nd^3)$
                                & \cite{hou2013basic} & {$dL+O(nd^2)$} & {$dL+O(nd^2)$} & $> {d^{4}n}L$ \\
& \cite{HouSCL16} & {$dL$} & {$dL$} &   {$\left(\frac{23}{16}d+\frac{5}{8}\right)dL$} \\
\bottomrule
\end{tabular}
\end{table}

In Section~\ref{sec::application}, we discuss how to extend our
decoding and repair schemes to some other MBR/MSR codes based
on the PM construction, so that these codes can gain certain
advantages we have for shift-XOR codes.  For codes based on the PM
construction in~\cite{rashmi2011optimal,HouSCL16}, the decoding and
repair schemes are the same except that the shift-XOR elimination is
replaced by certain sub-processes for finite fields and cyclic-shift
respectively.

% This paper is organized as follows.
% Section~\ref{sec::pre} formulates the shift-XOR regenerating codes, including both MBR and MSR codes, using the PM construction.
% Section~\ref{sec::mds} introduces our basic procedure of solving a system of shift-XOR equations, which will be used as a sub-process for decoding and repair of shift-XOR MBR and MSR codes using the PM construction.
% Section~\ref{sec::mbr} and Section~\ref{sec::msr} discuss the decoding and repair schemes for shift-XOR
% MBR and MSR codes, respectively.
% Section~\ref{sec::conclude} concludes this paper.

\section{Shift-XOR Storage Codes}
\label{sec::pre}

In this section, we formulate shift-XOR storage codes~\cite{sung2013zigzag,hou2013basic} after introducing some notations.

\subsection{Notations}

A range of integers from $i$ to $j$ is denoted by $\range{i}{j}$. When $i>j$, $\range{i}{j}$ is the empty set.
A (binary) sequence is denoted by a bold lowercase letter, e.g., $\mathbf{a}$.
The $i$-th entry of a sequence $\mathbf{a}$ is denoted by $\mathbf{a}[i]$.
The subsequence of $\mathbf{a}$ from the $i$-th entry to the $j$-th entry is denoted by $\mathbf{a}[\range{i}{j}]$.

For a sequence $\mathbf{a}$ of $L$ bits and a natural number
 $t$, the \emph{shift operator} $z^t$ pads $t$ zeros in front of $\mathbf{a}$, so that $z^t \mathbf{a} $ has $L+t$ bits and
\begin{equation*}
\left(z^t \mathbf{a} \right) \left[ l \right] =
\begin{cases}
0, & 1\le l \le t,\\
 \mathbf{a}  \left[ l-t \right], & t < l \le L+t.
\end{cases}
\end{equation*}
% Negative shifting means truncation, that is, $z^{-t} \mathbf{a}=\mathbf{a}[(t+1):L]$.
We use the convention that $\mathbf{a}[l]=0$ for $l\notin \range{1}{L}$, with which we can write
\begin{equation*}
  (z^t\mf a)[l]=\mathbf{a}[l-t], \quad l=1,\ldots,L+t.
\end{equation*}

Let $\mathbf{a}$ and $\mathbf{a}'$ be two sequences of length $L$ and $L'$, respectively.
The \emph{addition} of $\mathbf{a}$ and $\mathbf{a}'$, denoted by $\mathbf{a}+\mathbf{a}'$,  is a sequence of $\max\{L,L'\}$ bits obtained by bit-wise exclusive-or (XOR), i.e., for $l\in 1:\max\{L,L'\}$,
\begin{equation*}
  \left(\mathbf{a} + \mathbf{a}' \right) \left[ l \right] = \mathbf{a} \left[ l \right] \oplus \mathbf{a}' \left[ l \right],
\end{equation*}
where we also use the convention that $\mathbf{a}[l]=0$ for $l>L$ and $\mathbf{a}'[l]=0$ for $l>L'$.

\subsection{General Shift-XOR Storage Codes}
\label{sec:gen_enc}

We describe a general shift-XOR storage code and discuss special instances in the next subsection. 
The code has parameters $B$, $L$, $d$, $n$ and $\alpha$, which are positive integers. Consider a storage system of $n$ storage nodes employing a shift-XOR storage code.  
A message is formed by $B$ binary sequences, each of $L$ bits.
%, where the $i$-th message sequence is denoted as $\mathbf{x}_i$.
% \[
% \mathbf{x}_i = \left\{
% \mathbf{x}_i \left[1 \right], \mathbf{x}_i \left[2 \right], \ldots, \mathbf{x}_i \left[L \right]
% \right\},~~~~1 \le i \le L.
% \]
These $B$ message sequences are organized as a $d \times \alpha $ \emph{message matrix} $\mathbf{M}=(\mathbf{m}_{i,j})$ in certain way to be described subsequently, where two entries may share the same message sequence, and certain entries may be the all-zero sequences.
The \emph{generator matrix} used for encoding the message is an $n\times d$ matrix $\mathbf{\Psi} = (z^{t_{i,j}})$, where $t_{i,j}$ are nonnegative integers to be explained further soon.
For $1 \le i \le n, 1 \le j \le \alpha$, let
\[
{{\mathbf{y}}_{i,j}}=\sum\limits_{u=1}^{d}{{{z}^{{{t}_{i,u}}}}{{\mathbf{m}}_{u,j}}},
\]
called the \emph{coded sequence}.
Denote $\mathbf{Y}=(\mathbf{y}_{i,j})$ as the $n\times \alpha$ coded matrix of sequences $\mathbf{y}_{i,j}$, which can be written in the matrix form
%where $t_{i,u}$ are integers.
%Denote $\mathbf{\Psi} = (z^{t_{i,j}})$, called the generator matrix. 
%The above operation can be written in the matrix form
\begin{equation}
\label{equ::form}
\mathbf{Y}=\mathbf{\Psi}\mathbf{M}.
\end{equation}
%For $i=1,\ldots, n$, the $i$-th row of $\mathbf{Y}$ has the $\alpha $ sequences stored at the $i$-th storage node. 
% The $n\times \alpha$ coded sequences are obtained by~\eqref{equ::form}. As $\mathbf{\Psi}$ is of size $n\times d$ and $\mathbf{M}$ is of size $d\times \alpha$, $\mathbf{Y}$ is of size $n\times \alpha$.
  The $\alpha$ sequences in the $i$-th row of $\mathbf{Y}$ are stored at the $i$-th node (also called node $i$) of the storage system.

\begin{definition}[Refined Increasing Difference (RID) Property] \label{def:rip} Matrix $\mathbf{\Psi}=(z^{t_{i,j}})_{1\le i\le n, 1\le j\le d}$
  is said to satisfy the
  \emph{refined increasing difference (RID) property} if the following conditions hold: For any $i,{i}',j$,
    and ${j}'$ such that $i<{i}'$ and $j<{j}'$,
    \[0\leq {{t}_{i,{j}'}}-{{t}_{i,j}}<{{t}_{{i}',{j}'}}-{{t}_{{i}',j}},
    \]
    where equality in the first inequality holds only when $i=1$.
    We also say the numbers ${{t}_{i,j}}$, $1\le i\le n$, $1\le j\le d$ satisfy the RID property when they satisfy the above inequalities.
\end{definition}

To guarantee certain efficient decoding algorithms, in this paper, we require the generator matrix $\mathbf{\Psi}$ of a shift-XOR code satisfying the RID property. 
Different from the increasing difference property in~\cite{sung2013zigzag}, the RID property allows ${t}_{1,{j}'}-{t}_{1,j}=0$ such that less storage at each node is required. Suppose $\alpha=1$ and $B = d$, i.e., the $d$ entries of $\mathbf{M}$ are independent message sequences. The shift-XOR storage codes of this case have been studied in~\cite{sung2013zigzag}, and the zigzag algorithm can decode the $d$ message sequences from any $d$ rows of $\mathbf{Y}$.

Due to shift operations, the length of a coded sequence can be more than $L$ bits. In particular, the length of $\mathbf{y}_{i,j}$ is $L+t_{i,d}$. So the total number of bits stored at node $i$ is $\alpha (L+t_{i,d})$. The extra $\alpha t_{i,d}$ bits stored at node $i$ using a shift-XOR storage code is called the \emph{storage overhead}. Under the constraint of the RID property, it can be argued that the generator matrix minimizing the storage overhead is~\cite{Guo20}
\begin{equation*}
 \mathbf{\Psi}=(z^{(i-1)(j-1)}),   
\end{equation*}
which is a \emph{Vandermonde matrix}.
In our analysis, we suppose $t_{i,j} =O(nd)$, which is feasible as we have a choice of $t_{i,j}=(i-1)(j-1) < nd$.

%As the sequence length $L$ is independent of the other coding parameters $n,k,d,\alpha$, the storage overhead can be marginalized using a relative large $L$.
%In a practical storage system, $L$ can be thousands or even larger.
% Though more bits are stored, the shift-XOR codes  lower communication and computation costs than cyclic-shift-and-XOR and finite-field based codes, such as the algorithms to be discussed in this paper. 

% and hence each node stores more than $\alpha L$ bits. More precisely, bits are , i.e., the storage overheads at node $i$ are , which are the same as in~\cite{hou2013basic}. The storage overheads of the cyclic-shift PM codes in~\cite{HouSCL16} are $\alpha$ bits. And there is no storage overhead for finite-field PM codes~\cite{rashmi2011optimal}.
% Without the loss of generality, all entries in the first column of $\mathbf{\Psi}$ are assumed to be $z^0$.

\subsection{Shift-XOR Regenerating Codes}
\label{sec:shift-xor-reg}

We discuss two classes of shift-XOR product-matrix (PM) regenerating
codes~\cite{hou2013basic}. The two constructions of the message matrix
$\mathbf{M}$ are the same as those of the (finite-field) PM MBR and
MSR codes~\cite{rashmi2011optimal}. According to the storage overhead discussed in the preceding section,
these codes achieve the MBR/MSR tradeoff asymptotically when
$L\rightarrow\infty$, so the constructed codes are
called \emph{shift-XOR MBR} codes and \emph{shift-XOR MSR} codes,
respectively.

In contrast, regenerating codes using cyclic-shift-and-XOR
operations~\cite{HouSCL16} and finite-field
operations~\cite{rashmi2011optimal} do not have storage
overhead. Though with the storage overhead, the shift-XOR codes have
the potential of low encoding/decoding complexity, to be demonstrated
by the schemes of this paper.

\subsubsection{Shift-XOR MBR Codes}
\label{sec:constr-xor-mbr}

% Minimum Bandwidth Regenerating (MBR) codes achieve the optimal repair bandwidth. To construct a MBR code with
Fix an integer $k$ with $k \leq d$. Consider $\alpha =d$ and
\begin{equation}
  \label{eq:xormbr}
  B
 = \frac{1}{2} (k+1)k+ k(d-k).
% = kd- \binom{k}{2}.
\end{equation}
The message matrix $\mathbf{M} $ is of the form
\begin{equation}
\label{eqn:message_matrix}
 \mathbf{M}=
 \begin{bmatrix}
 \mathbf{S} & \mathbf{T}\\
 \mathbf{T}^\top & \mathbf{O}
 \end{bmatrix},
\end{equation}
where $\mathbf{S}$ is a $k\times k$ symmetric matrix of the first $\frac{1}{2} \left(k+1 \right)k$ message sequences,
$\mathbf{T}$ is a $k \times \left(d-k\right)$ matrix of the remaining $k \left(d-k\right)$ message sequences,
 $\mathbf{O}$ is a $\left(d-k\right)\times \left(d-k\right)$ matrix of the zero sequence $\mathbf{0}$, and $\mathbf{T}^\top$ is the transpose of $\mathbf{T}$.
 By \eqref{eq:xormbr}, all the $B$ message sequences are used in $\mathbf{M}$.

The shift-XOR MBR code has the coded sequences $\mathbf{Y}=\mathbf{\Psi}\mathbf{M}$, where $\mathbf{\Psi} = (z^{t_{i,j}})$ is an $n\times d$ matrix satisfying the {RID property}.
%for any $i,{i}',j$,
% and ${j}'$ such that $i<{i}'$ and $j<{j}'$,
%	\[0\leq {{t}_{i,{j}'}}-{{t}_{i,j}}<{{t}_{{i}',{j}'}}-{{t}_{{i}',j}},	\]
%where the first equality holds only when $i=1$.
%Differently from the increasing difference property~\cite{sung2013zigzag}, the RID property can accept ${t}_{1,{j}'}-{t}_{1,j}=0$ such that less storage at each node is required.
%Without the loss of generality, all entries in the first column of $\mathbf{\Psi}$ are assumed to be $z^0$.
A shift-XOR MBR code has the parameters $n$, $k$, $d$ and $L$, and is usually referred to as an $[n,k,d]$ code.

\begin{example}[$[6,3,4 \rbracket$ shift-XOR MBR Code]
\label{exam::mbr}
For a shift-XOR MBR code with $n=6$, $k=3$, $d=4$, the message matrix is of the form
\[
\mathbf{\mathbf{M} }=\begin{bmatrix}
   {\mathbf{x}_{1}} & {\mathbf{x}_{2}}   & {\mathbf{x}_{3}} & {\mathbf{x}_{7}}\\
   {\mathbf{x}_{2}} & {\mathbf{x}_{4}} & {\mathbf{x}_{5}} & {\mathbf{x}_{8}}\\
   {\mathbf{x}_{3}} & {\mathbf{x}_{5}}  & {\mathbf{x}_{6}} & {\mathbf{x}_{9}} \\
   {\mathbf{x}_{7}} & {\mathbf{x}_{8}}  & {\mathbf{x}_{9}} & {\mathbf{0}}
\end{bmatrix},
\]
where $\mathbf{x}_i$ is a binary sequence of $L$ bits.
Using the Vandermonde generator matrix %$\mathbf{\Psi}$ is in the form of
\[
\mathbf{\Psi }=\begin{bmatrix}
   {1} & {1} & {1}  & {1}  \\
   {1} & {{z}} & {{z}^{2}} & {{z}^{3}}  \\
   {1} & {{z^{2}}} & {{z}^{4}} & {{z}^{6}}  \\
   {1} & {{z^{3}}} & {{z}^{6}} & {{z}^{9}}  \\
   {1} & {{z^{4}}} & {{z}^{8}} & {{z}^{12}}  \\
   {1} & {{z^{5}}} & {{z}^{10}} & {{z}^{15}}  \\
\end{bmatrix},
\]
the coded sequences stored at node $i$ ($1 \le i \le 6$) are
\begin{equation*}
\left\{
\begin{IEEEeqnarraybox}[][c]{r.C.l}
\mathbf{y}_{i,1} \!&\!=\!&\! \mathbf{x}_{1}+z^{i-1}\mathbf{x}_{2}+z^{2(i-1)}\mathbf{x}_{3}+z^{3(i-1)}\mathbf{x}_{7},  \\
\mathbf{y}_{i,2} \!&\!=\!&\! \mathbf{x}_{2}+z^{i-1}\mathbf{x}_{4}+z^{2(i-1)}\mathbf{x}_{5}+z^{3(i-1)}\mathbf{x}_{8},  \\
\mathbf{y}_{i,3} \!&\!=\!&\! \mathbf{x}_{3}+z^{i-1}\mathbf{x}_{5}+z^{2(i-1)}\mathbf{x}_{6}+z^{3(i-1)}\mathbf{x}_{9},  \\
\mathbf{y}_{i,4} \!&\!=\!&\! \mathbf{x}_{7}+z^{i-1}\mathbf{x}_{8}+z^{2(i-1)}\mathbf{x}_{9}.
\end{IEEEeqnarraybox}\right.
\end{equation*}
\end{example}

\subsubsection{Shift-XOR MSR Codes}
\label{sec::cons-xor-msr}
Here we only discuss shift-XOR MSR codes with $d=2k-2$ and $\alpha=k-1$.
Codes with $d\geq 2k-1$ can be constructed using the method in~\cite{rashmi2011optimal} based on the codes with $d=2k-2$ and $\alpha=k-1$. Let $B = k\alpha=(\alpha+1)\alpha$. The message matrix $\mathbf{M}$ is of the form
\begin{equation} \label{eq:msr-m}
\mathbf{M}=
  \begin{bmatrix}
    \mathbf{S}\\
    \mathbf{T}
  \end{bmatrix},
\end{equation}
where ${{\mathbf{S}}}$ is an $\alpha \times \alpha $ symmetric matrix of the first $\frac{1}{2}\alpha \left( \alpha +1 \right)$ message sequences, and  ${{\mathbf{T}}}$ is another $\alpha\times \alpha$ symmetric matrix of the remaining  $\frac{1}{2}\alpha \left( \alpha +1 \right)$ message sequences.

The generator matrix $\mathbf{\Psi}$ is of the form 
\begin{equation}\label{eq:msr:p}
\mathbf{\Psi}=
\begin{bmatrix}
  \mathbf{\Phi} & \mathbf{\Lambda\Phi}
\end{bmatrix},
\end{equation}
where $\mathbf{\Phi}= (z^{{t}_{i,j}})$ is an $n \times \alpha$ matrix satisfying the RID property, and
$\mathbf{\Lambda}$ is an $n\times n$ diagonal matrix with diagonal entries $z^{{\lambda}_{1}},z^{{\lambda}_{2}},\ldots,z^{{\lambda}_{n}}$ such that
$\mathbf{\Psi}$ satisfies the RID property.
When $\mathbf{\Phi}=\left(z^{(i-1)(j-1)}\right)$ and ${\lambda}_i=(i-1)\alpha$, $\mathbf{\Psi}$ is  a Vandermonde matrix, for which the storage overheads are minimal as we have discussed.
% To simplify our results on MSR codes, we often assume that the elements in the first column of $\mathbf{\Phi}$ are $z^{0}$, which does not affect the transmission bandwidth and decoding complexity.

The shift-XOR MSR code described above has the coded sequences $\mathbf{Y} =  \mathbf{\Psi}\mathbf{M}$, and is usually referred to as an $[n,k,d]$ code.

\begin{example}[$[6,3,4 \rbracket$ shift-XOR MSR Code]
\label{exam::msr}
For a shift-XOR MSR code with $n=6$, $k=3$, $d=4$ and $\alpha=k-1=2$, the message matrix is
\begin{equation*}
\mathbf{M}= \begin{bmatrix}
   {\mathbf{x}_{1}} & {\mathbf{x}_{2}}   \\
   {\mathbf{x}_{2}} & {\mathbf{x}_{3}} \\
   {\mathbf{x}_{4}} & {\mathbf{x}_{5}}   \\
   {\mathbf{x}_{5}} & {\mathbf{x}_{6}}
\end{bmatrix},
\end{equation*}
where $\mathbf{x}_i$ is a binary sequence of $L$ bits.
Let
\[
\mathbf{\Phi }=\begin{bmatrix}
   {1} & {1}   \\
   {1} & {{z}}  \\
   {1} & {{z^{2}}}   \\
   {1} & {{z^{3}}}  \\
   {1} & {{z^{4}}}  \\
   {1} & {{z^{5}}}  \\
\end{bmatrix}.
\]
and $\mathbf{\Lambda}= \text{diag} \left\{1,z^{2},z^{4},z^{6},z^{8},z^{10}\right\}$.
Then the generator matrix $\mathbf{\Psi}$ is
\[
\mathbf{\Psi }=\begin{bmatrix}
   {1} & {1} & {1}  & {1}  \\
   {1} & {{z}} & {{z}^{2}} & {{z}^{3}}  \\
   {1} & {{z^{2}}} & {{z}^{4}} & {{z}^{6}}  \\
   {1} & {{z^{3}}} & {{z}^{6}} & {{z}^{9}}  \\
   {1} & {{z^{4}}} & {{z}^{8}} & {{z}^{12}}  \\
   {1} & {{z^{5}}} & {{z}^{10}} & {{z}^{15}}  \\
\end{bmatrix}.
\]
The coded sequences stored at node $i$ ($1 \le i \le 6$) are
\begin{equation*}
  \left\{
\begin{IEEEeqnarraybox}[][c]{r.C.l}
\mathbf{y}_{i,1} \!&\!=\!&\! \mathbf{x}_{1}+z^{i-1}\mathbf{x}_{2}+z^{2 \left(i-1\right)}\mathbf{x}_{4}+z^{3\left(i-1\right)}\mathbf{x}_{5},  \\
\mathbf{y}_{i,2} \!&\!=\!&\! \mathbf{x}_{2}+z^{i-1}\mathbf{x}_{3}+z^{2\left(i-1\right)}\mathbf{x}_{5}+z^{3\left(i-1\right)}\mathbf{x}_{6}.
\end{IEEEeqnarraybox}\right.
\end{equation*}
\end{example}

\section{Solving a System of Shift-XOR Equations}
\label{sec::mds}
Before introducing the decoding and repair schemes of the shift-XOR regenerating codes, we give an algorithm for solving a system of shift-XOR equations, called \textit{shift-XOR elimination}. This algorithm will be used as a sub-process of our subsequent decoding and repair schemes, and is of independent interest due to its fundamental role.

A $k\times k$ system of shift-XOR equations is given by
\begin{equation}
\label{equ::mds_enc}
\begin{bmatrix}
   {\mathbf{y}_{1}}  \\
   {\mathbf{y}_{2}}  \\
   {\vdots} \\
   {\mathbf{y}_{k}}  \\
\end{bmatrix}=\mathbf{\Psi }
\begin{bmatrix}
   {\mathbf{x}_{1}}  \\
   {\mathbf{x}_{2}}  \\
   {\vdots} \\
   {\mathbf{x}_{k}}  \\
\end{bmatrix},
\end{equation}
where ${{\mathbf{x}}_{1}},{{\mathbf{x}}_{2}},\ldots ,{{\mathbf{x}}_{k}}$ are binary sequences of $L$ bits, and matrix $\mathbf{\Psi }=(z^{t_{i,j}})$ satisfies the RID property given in Definition~\ref{def:rip}. The problem of solving a system of shift-XOR equations~\eqref{equ::mds_enc} is to calculate $\mathbf{x}_i$, $i=1,\ldots,k$ for given $\mathbf{y}_{i}$, $i=1,\ldots,k$ and $\mathbf{\Psi}$.

% Let $k>0$ be an integer.
% A system of shift-XOR equations has a $k\times k$ generator  Let
% and let $\mathbf{y}_{1},\mathbf{y}_{2},\ldots,\mathbf{y}_{k}$ be sequences generated by 

\subsection{Zigzag Algorithm}
\label{sec::zigzag}

One approach to solve the system is the zigzag algorithm~\cite{sung2013zigzag}, which performs successive cancellation. We use an example to illustrate the idea of the zigzag algorithm.

\begin{example}
Consider the $3\times 3$ system
\begin{equation} \label{eq:sys:ex}
\begin{bmatrix}
\mathbf{y}_1\\
\mathbf{y}_2\\
\mathbf{y}_3
\end{bmatrix}
=\begin{bmatrix}
1 & {z} & {z^2}\\
1 & {z^2} & {z^4}\\
1 & {z^3} & {z^6}
\end{bmatrix}
\begin{bmatrix}
\mathbf{x}_1\\
\mathbf{x}_2\\
\mathbf{x}_3
\end{bmatrix}.
\end{equation}
Table~\ref{tab:ex} illustrates how the bits in $\mathbf{y}_i$ are aligned with bits in $\mathbf{x}_1$, $\mathbf{x}_2$ and $\mathbf{x}_3$. We see that $\bx_1[1]=\by_i[1]$, for $i=1,2,3$, and hence $\bx_1[1]$ is solvable. Next, we see that $\bx_1[2]=\by_i[2]$, for $i=2,3$ and hence $\bx_1[2]$ is solvable. Substituting $\bx_1[2]$ back into $\by_1$, we further obtain $\bx_2[1]=\by_1[2]-\bx_1[2]$.
This process can be repeated to solve all the bits in $\bx_1$, $\bx_2$ and $\bx_3$: in every iteration, a solvable bit is found and is substituted back to all the equations it involves in. When there are more than one solvable bits in an iteration, one of them is chosen for substitution. 
% Substituting $\bx_1[1:3]$ back into $\by_1$ and $\by_2$, we can further obtain $\bx_2[1]=\by_2[3]-\bx_1[3]$.
% Substituting $\bx_2[1]$ back into $\by_2$ and $\by_3$, we get $\bx_1[4] = \by_3[4]-\bx_2[1]$.
\end{example}

%Table~\ref{tab:ex} illustrates how the bits in $\mathbf{y}_i$ are aligned with bits in $\mathbf{x}_1$, $\mathbf{x}_2$ and $\mathbf{x}_3$. We see that $\bx_1[1]=\by_1[1]$, $\bx_1[2]=\by_1[2]$ and $\bx_1[3]=\by_3[3]$. Substituting $\bx_1[1:3]$ back into $\by_1$, $\by_2$ and $\by_3$, we can further obtain $\bx_2[1]=\by_1[2]-\bx_1[2]$. Substituting $\bx_2[1]$ back into $\by_1$, $\by_1$ and $\by_3$, we get $\bx_1[4] = \by_3[4]-\bx_2[1]$. This process can be repeated until all the bits in $\bx_1$, $\bx_2$ and $\bx_3$ are decoded.

\begin{table}
  \centering
  \caption{The three tables illustrate how $\mathbf{y}_1, \by_2,\by_3$ are formed by $\mathbf{x}_1$, $\mathbf{x}_2$ and $\mathbf{x}_3$. For number $l$ in the row of $\mathbf{y}_i$, let $l_j$ be the number in the same column of $l$ and in the row of $\bx_j$. Then the table tells that $\mathbf{y}_i[l] = \bx_1[l_1]+\bx_2[l_2]+\bx_3[l_3]$. For example, the 2nd, 3rd and 4th columns of Table \ref{tab:1:1} mean $\by_1[1]=\bx_1[1]$, $\by_1[2] = \bx_1[2]+\bx_2[1]$ and $\by_1[3] = \bx_1[3]+\bx_2[2]+\bx_3[1]$, respectively.} \label{tab:ex}
  \subfloat[][$\mathbf{y}_1 = \mathbf{x}_1 + z\mathbf{x}_2 + z^2\mathbf{x}_3$]{\label{tab:1:1}
  \begin{tabular}[ht]{c|ccccccccc}
    \toprule 
    $\mathbf{y}_1$ & 1 & 2 & 3 & 4 & 5 & $\cdots$ & $L$ & $L+1$ & $L+2$ \\
    \hline
    $\mathbf{x}_1$ & 1 & 2 & 3 & 4 & 5 & $\cdots$ & $L$ & & \\
    $\mathbf{x}_2$ & & 1 & 2 & 3 & 4 & $\cdots$ & $L-1$ & $L$ & \\
    $\mathbf{x}_3$ & & & 1 & 2 & 3 & $\cdots$ & $L-2$ & $L-1$ & $L$ \\
    % \hline
    % $\hat\bx_3$ & & & 1 & 2 & 3 & $\cdots$ & $L-2$ & $L-1$ & $L$  \\
    \bottomrule
  \end{tabular}}

  \subfloat[][$\mathbf{y}_2 = \mathbf{x}_1 + z^2\mathbf{x}_2 + z^4\mathbf{x}_3$]{
  \begin{tabular}[ht]{c|ccccccccccc}
    \toprule 
    $\mathbf{y}_2$ & 1 & 2 & 3 & 4 & 5 & $\cdots$ & $L$ & $L+1$ & $L+2$ & $L+3$ & $L+4$ \\
    \hline
    $\mathbf{x}_1$ & 1 & 2 & 3 & 4 & 5 & $\cdots$ & $L$ & & \\
    $\mathbf{x}_2$ & & & 1 & 2 & 3 & $\cdots$ & $L-2$ & $L-1$ & $L$ \\
    $\mathbf{x}_3$ & & & & & 1 & $\cdots$ & $L-4$ & $L-3$ & $L-2$ & $L-1$ & $L$ \\
    % \hline
    % $\hat\bx_2$ & & & 1 & 2 & 3 & $\cdots$ & $L-2$ & $L-1$ & $L$ \\
    \bottomrule
  \end{tabular}}

    \subfloat[][$\mathbf{y}_3 = \mathbf{x}_1 + z^3\mathbf{x}_2 + z^6\mathbf{x}_3$]{
  \begin{tabular}[ht]{c|ccccccccccccc}
    \toprule 
    $\mathbf{y}_3$ & 1 & 2 & 3 & 4 & 5 & 6 & 7 & $\cdots$ & $L$ & $L+1$ & $L+2$ & $L+3$ & $\cdots$ \\
    \hline
    $\mathbf{x}_1$ & 1 & 2 & 3 & 4 & 5 & 6 & 7 & $\cdots$ & $L$ & & \\
    $\mathbf{x}_2$ & & & & 1 & 2 & 3 & 4 & $\cdots$ & $L-3$ & $L-2$ & $L-1$ & $L$ \\
    $\mathbf{x}_3$ & & & & & & & 1 & $\cdots$ & $L-6$ & $L-5$ & $L-4$ & $L-3$ & $\cdots$ \\
    % \hline
    % $\hat\bx_1$ & 1 & 2 & 3 & 4 & 5 & 6 & 7 & $\cdots$ & $L$ & & \\
    \bottomrule
  \end{tabular}}
\end{table}

The zigzag algorithm in~\cite{sung2013zigzag} implements the above idea to solve any $k\times k$ system of shift-XOR equations as defined in \eqref{equ::mds_enc}.
The zigzag algorithm, however, is not optimal in several aspects. 
First, the zigzag algorithm needs all the $kL+\sum_{i=1}^kt_{i,k}$ bits of $\mathbf{y}_{1},\mathbf{y}_{2},\ldots,\mathbf{y}_{k}$ to solve the $kL$ bits of ${{\mathbf{x}}_{1}},{{\mathbf{x}}_{2}},\ldots, {{\mathbf{x}}_{k}}$. The extra $\sum_{i=1}^kt_{i,k}$ bits consumed by the algorithm is called the \emph{communication overhead} or \emph{bandwidth overhead}. The ideal case is to have zero bandwidth overhead, same as solving a full-rank $k\times k$ system of linear equations over a finite field.

Second, the space and time computation complexities of the zigzag algorithm have room to improve.
The zigzag algorithm~\cite{sung2013zigzag} runs in $kL$ iterations to solve all the $kL$ bits. In each iteration, a solvable bit is found to back substitute into all the related equations. In~\cite{sung2013zigzag}, two approaches for searching the solvable bit are discussed. The first approach takes $O(k^2)$ comparisons to find a solvable bit and results in totally $O(k^3L)$ time complexity. To reduce the time complexity, the second approach uses a pre-calculated array of $O(kL)$ integers to assist the searching process. 
Searching and updating the array takes $O(k)$ integer operations in each iteration, so that the time complexity is $O(k^2L)$.
In addition to the input and output sequences, the second approach requires $O(kL)$ auxiliary space to store the integer array.
Without otherwise specified, we refer to the second approach as the zigzag algorithm.

\subsection{Shift-XOR Elimination}

Here we propose an algorithm to solve systems of shift-XOR equations, 
called \emph{shift-XOR elimination}, which improves the zigzag algorithm in terms of both bandwidth overhead and computation complexities.
First, only a subsequence of $L$ bits of $\by_i$ ($i=1,\ldots,k$) is used so that the shift-XOR elimination has no bandwidth overhead. Second, the order of the bits to solve follows a regular pattern so that the shift-XOR elimination has lower 
computation time and space costs than the zigzag algorithm. We use an example to illustrate our algorithm.

\begin{example}
  Consider the system in \eqref{eq:sys:ex}. As illustrated in Table~\ref{tab:ex}, $\by_1[1]$, $\by_2[1]$ and $\by_3[1]$ are all equal to $\bx_1[1]$ and hence one of them is sufficient for solving $\bx_1[1]$ and other two are redundant. Similarly, $\by_2[2]$ and $\by_3[2]$ are the same, and one of them is redundant.
Define subsequences
\begin{IEEEeqnarray*}{rCl}
  \hat\bx_1 & = & \by_3[1:L], \\
  \hat\bx_2 & = & \by_2[3:(L+2)], \\
  \hat\bx_3 & = & \by_1[3:(L+2)].
\end{IEEEeqnarray*}
Table~\ref{tab:ex:dec} illustrates how $\hat\bx_i$ is formed by $\bx_1, \bx_2, \bx_3$.
In particular, for $l\in 1:L$,
\begin{IEEEeqnarray*}{rCl}
  \hat\bx_1[l] & = & \bx_1[l] + \bx_2[l-3] + \bx_3[l-6], \\
  \hat\bx_2[l] & = & \bx_2[l] + \bx_1[l+2] + \bx_3[l-2], \\
  \hat\bx_3[l] & = & \bx_3[l] + \bx_1[l+2] + \bx_2[l+1].
\end{IEEEeqnarray*}
We see that $\hat\bx_i$ involves all the bits in $\bx_i$.

Let's illustrate how to solve the system using $\hat\bx_i$, $i=1,2,3$. 
The system is solved in multiple iterations indexed by $\iters=1,2,\ldots$, which can be further separated into three phases:
\begin{enumerate}
\item For each iteration $\iters=1,2$, one bit in $\bx_1$ is solved.
\item For the iteration $\iters=3$, one bit is solved in $\bx_1$ and one bit is solved in $\bx_2$ sequentially.
\item For each iteration $\iters \geq 4$, one bit is solved from each of $\bx_1, \bx_2$ and $\bx_3$ sequentially. (When $l>L$, $\bx_i[l]$ is supposed to be solvable.)
\end{enumerate}
Here, a bit is solved implies that it is also back substituted into the equations it involves. 
Table~\ref{tab:ex:dec} illustrates this order of bit solving. % where we use $0\leq l_i \leq L$ ($i=1,2,3$) to indicate the number of bits solved in $\bx_i$ after an iteration. Initially, $l_1=l_2=l_3=0$.
The first two iterations form the first phase, where $\bx_1[1]$ and $\bx_1[2]$ are solved. The third iteration forms the second phase, where $\bx_1[3]$ is solved first and then $\bx_2[1]$ is solved by substituting $\bx_1[3]$. Other iterations form the third phase. At iteration $4$, for example, $\bx_1[4]$ is first solved by substituting $\bx_2[1]$; $\bx_2[2]$ is then solved by substituting $\bx_1[4]$; last, $\bx_3[1]$ is solved by substituting $\bx_1[3]$ and $\bx_2[2]$.
% If $l_1 < \min\{l_2+3,l_3+6\}$, we can solve $\bx_1[l_1+1]$ by
% \begin{equation*}
%   \bx_1[l_1+1] = \hat\bx_1[l_1+1] + \bx_2[l_1-2] + \bx_3[l_1-5],
% \end{equation*}
% where $\bx_2[l_1-2]$ has been solved as $l_1-2 \leq l_2$ and $\bx_3[l_1-5]$ has been solved as $l_1-5 \leq l_3$. Similarly, if $l_2 < \min\{l_1-2,l_3+2\}$, then $\bx_2[l_2+1]$ can be solved; if $l_3 < \min\{l_1-2,l_2-1\}$, then $\bx_3[l_3+1]$ can be solved.
\end{example}

\begin{table}
  \centering
  \caption{Solving system \eqref{eq:sys:ex} by the Shift-XOR Elimination for the first $10$ iterations. The first row gives the three phases of the iterations, and the second row gives the iterations $\iters$. % The row $l_i$ ($i=1,2,3$) says how many bits are solved in $\bx_i$ after an iteration.
    The three rows following $\hat\bx_i$ ($i=1,2,3$) show how $\hat\bx_i$ is formed by $\bx_j$, $j=1,2,3$. For number $l$ in the row of $\hat\bx_i$, let $l_j$ be the number in the same column of $l$ and in the row of $\bx_j$ following $\hat\bx_i$. Then the table tells that $\hat\bx_i[l] = \bx_1[l_1]+\bx_2[l_2]+\bx_3[l_3]$. For each iteration, the bits decoded are specified by the  entries in the same column and in the gray rows. 
    For example, from the column indexed by $\iters=4$, the three gray entries are $4$, $2$ and $1$, where the entry $4$ in the row of $\hat\bx_1$ means that $\bx_1[4]$ can be solved by substituting the previous solved bits into $\hat\bx_1[4]$.} \label{tab:ex:dec}
%  \begin{tabular}[th]{c|cc|cc|ccc|ccc|ccc}
%    \toprule
%    iteration & 1 & 2 & \multicolumn{2}{c|}{3} &  \multicolumn{3}{c|}{4} & \multicolumn{3}{c|}{5} & \multicolumn{3}{c}{6} \\
%    \hline
%    index & 1 & 1 & 1 & 2 & 1 & 2 & 3 & 1 & 2 & 3 & 1 & 2 & 3 \\
%    $l_1$ & 1 & 2 & 3 & 3 & 4 & 4 & 4 & 5 & 5 & 5 & 6 & 6 & 6 \\
%    $l_2$ & 0 & 0 & 0 & 1 & 1 & 2 & 2 & 2 & 3 & 3 & 3 & 4 & 4 \\
%    $l_3$ & 0 & 0 & 0 & 0 & 0 & 0 & 1 & 1 & 1 & 2 & 2 & 2 & 3 \\
%    \bottomrule
%  \end{tabular}
 \begin{tabular}[ht]{c|cc|c|ccccccccccc}
   \toprule
   {phase} & \multicolumn{2}{c|}{1} & 2 & \multicolumn{8}{c}{3} \\
   \hline
   iteration $\iters$ & 1 & 2 & 3 & 4 & 5 & 6 & 7 & 8 & 9 & 10 & $\cdots$ \\
 %   \hline
 % {$l_1$} & 1 & 2 & 3 & 4 & 5 & 6 & 7 & 8 & 9 & 10 & $\cdots$ \\
 % {$l_2$} & 0 & 0 & 1 & 2 & 3 & 4 & 5 & 6 & 7 & 8  & $\cdots$ \\
 % {$l_3$} & 0 & 0 & 0 & 1 & 2 & 3 & 4 & 5 & 6 & 7  & $\cdots$ \\
    \hline
   \rowcolor{lightgray}   $\hat\bx_1$ & 1 & 2 & 3 & 4 & 5 & 6 & 7 & 8 & 9 & 10 & $\cdots$ \\
       $\bx_1$ & 1 & 2 & 3 & 4 & 5 & 6 & 7 & 8 & 9 & 10 & $\cdots$ \\
        $\bx_2$  & & & & {1} & {2} & {3} & {4} & {5} & {6} & {7} &  $\cdots$ \\
       $\bx_3$ & & & & & & & {1} & {2} & {3} & {4} &  $\cdots$ \\
   \hline
   \rowcolor{lightgray}    $\hat\bx_2$   & & & 1 & 2 & 3 & 4 & 5 & 6 & 7 & 8 & $\cdots$\\
     $\bx_1$ &  &  & {3} & {4} & {5} & {6} & {7} & {8} & {9} & {10} & $\cdots$\\
     $\bx_2$ & & & 1 & 2 & 3 & 4 & 5 & 6 & 7 & 8 & $\cdots$\\
     $\bx_3$ & & & & & {1} & {2} & {3} & {4} & {5} & {6} & $\cdots$\\
   \hline
   \rowcolor{lightgray}    $\hat\bx_3$ & & & & 1 & 2 & 3 & 4 & 5 & 6 & 7 & $\cdots$\\
      $\bx_1$ & &  &   & {3} & {4} & {5} & {6} & {7} & {8} & {9} & $\cdots$\\
      $\bx_2$ & & &  & {2} & {3} & {4} & {5} & {6} & {7} & {8} & $\cdots$\\
 $\bx_3$ & & & & 1 & 2 & 3 & 4 & 5 & 6 & 7 & $\cdots$\\
    \bottomrule
  \end{tabular}
\end{table}

Now we introduce the general shift-XOR elimination for solving~\eqref{equ::mds_enc}.
The algorithm uses the subsequence $\hat{\mathbf{x}}_i$, $i=1,\ldots,k$ defined as
\begin{equation}\label{eq:xhat}
  {{\hat\bx}_i} = \mathbf{y}_{k+1-i}\left[ t_{k+1-i,i}+(\range{1}L) \right],
\end{equation}
where $t_{k+1-i,i}+(\range{1}L)$ denotes $\range{(t_{k+1-i,i}+1)}{(t_{k+1-i,i}+L)}$.
% Instead of $\mathbf{y}_{1},\mathbf{y}_{2},\ldots,\mathbf{y}_{k}$,
As $\hat{\mathbf{x}}_i$ has exactly $L$ bits, our algorithm needs exactly $kL$ input bits and hence achieves zero bandwidth overhead.
Substituting \eqref{equ::mds_enc} into \eqref{eq:xhat}, we have for $1\leq l \leq L$,
\begin{equation}\label{eq:xhat2}
  \hat\bx_i[l] = \bx_i[l] + \sum_{j\neq i} \bx_j[l-t_{k+1-i,j} + t_{k+1-i,i}],
\end{equation}
where we use the convention that $\bx_i[l]=0$ for $l\leq 0$ and $l>L$.

The shift-XOR elimination solves $\hat{\mathbf{x}}_i$, $i=1,\ldots,k$ as follows. The algorithm runs in a number of iterations indexed by $\iters=1,2\ldots$, which are partitioned into $k$ phases. For $b=1,2,\ldots,k$, define
\begin{equation} 
\label{eq:tb}
L_{b}=
\begin{cases}
t_{k-b,b+1}-t_{k-b,b}, & 1\leq b <k, \\
L, & b=k,
\end{cases}
\end{equation}
and define $L_{1:b} = \sum_{b'=1}^bL_{b'}$.
The $b$-th phase ($b\in 1:k$) has $L_b$ iterations. 
The operations in each iteration are specified as follows:
\begin{itemize}
\item For each iteration $\iters$ in phase $1$ (i.e., $\iters\in 1:L_1$), $\bx_1[\iters]$ is solved (using $\hat\bx_1[\iters]$). 
\item For each iteration $\iters$ in phase $b=2,\ldots, k$ (i.e., $\iters\in L_{1:(b-1)}+(1:L_{b})$), $\bx_i[\iters-L_{1:(i-1)}]$ is solved (using $\hat\bx_i[\iters-L_{1:(i-1)}]$ and the previously solved bits) sequentially for $i=1,\ldots,b$. 
%\item For each iteration $l$ in $\sum_{b'=1}^{k-1}L_{b'}+(1:L)$, $\bx_i[l-\sum_{b'=1}^{i-1}L_{b'}]$ is solved (using $\hat\bx_i[l-\sum_{b'=1}^{i-1}L_{b'}]$ and the previously solved bits) sequentially for $i=1,\ldots,k$.
\end{itemize}
In the above process, i) a bit is solved implies that it is also back substituted into the equations it involves in; ii) for $\bx_i[l]$ with $l>L$, $\bx_i[l]$ is supposed to be solvable as $0$, and hence do not need to be substituted; iii) the total number of iterations is $L_{1:k}$.

A pseudocode of the shift-XOR elimination is given in Algorithm~\ref{alg::mds}, which also demonstrates an \emph{in-place} implementation of the shift-XOR elimination. % To simplify the calculation of the index of the bit to solve, we use $l_i$ to denote the number of bits solved in $\bx_i$ during the solving process.
The loop started at Line \ref{alg:sx:sg1} enumerates all the phases $b=1,\ldots,k$. The operations in each iteration are given from Line \ref{alg:sx:op1} to \ref{alg:sx:op2}: In Line \ref{alg:sx:dec}, one more bit is marked to be solved, and the following three lines perform back substitution. For in-place implementation, the back substitution result $\hat{\bx}_{v}[l+t_{{k+1-v},i}-t_{{k+1-v},v}]\oplus \bx_i[ l]$ is stored at the same place of $\hat{\bx}_{v}[l+t_{{k+1-v},i}-t_{{k+1-v},v}]$. After the execution of the algorithm, the value  $\bx_i[l]$ is stored at the same storage space as $\hat\bx_i[l]$.

\begin{algorithm}
\caption{Shift-XOR elimination with in-place implementation. After the execution, the value $\bx_i[l]$ is stored at the same storage space as $\hat\bx_i[l]$. 
}
\label{alg::mds}
\begin{algorithmic}[1]
\REQUIRE sequences ${{\hat\bx}_i}$, $1\le i\le k$.
\ENSURE solved sequences ${\bx_i}$, $1\le i\le k$.

\STATE {Initialize $\iters \leftarrow 0$}

\FOR {$b\leftarrow 1:k$} \label{alg:sx:sg1}
  \FOR { $L_{b}$ iterations }\label{alg:sx:op1}
  \STATE $\iters \leftarrow \iters+1$; 
  \FOR { $i\leftarrow 1:b$ }
  \STATE $l \leftarrow \iters-L_{1:(i-1)}$; (the value of $\bx_i[l]$ is stored at the same place of $\hat\bx_i[l]$) \label{alg:sx:dec}
  \FOR {$j\leftarrow 1,2,\ldots,i-1,i+1,\ldots,k$}
  \IF {$0<{l}+{t_{{k+1-j},i}}-{t_{{k+1-j},j}}\le L$ }
  \STATE $\hat{\bx}_{j}[l+t_{{k+1-j},i}-t_{{k+1-j},j}]\oplus \leftarrow \bx_i[ l]$; (in-place back substitution) \label{alg:sx:op2}
  \ENDIF
  \ENDFOR
  \ENDFOR
  \ENDFOR
  \ENDFOR
\end{algorithmic}
\end{algorithm}

To prove the correctness of the shift-XOR elimination, we only need to show that each bit chosen to solve during the execution of the algorithm can be expressed as $\hat\bx_i$, $i=1,\ldots,k$ and the previously solved bits.
Theorem~\ref{theo::mds}, proved in Appendix, justifies the shift-XOR elimination.

\begin{theorem}
\label{theo::mds}
Consider a $k\times k$ system of shift-XOR equations
$(\by_1 \ \cdots \ \by_k)^\top 
= \mathbf{\Psi}
(\bx_1 \ \cdots \ \bx_k)^\top
$ with $\mathbf{\Psi}$ satisfying the RID property.
The shift-XOR elimination can successfully solve $\bx_i$, $i=1,\ldots,k$ using 
\begin{equation*}
  {{\hat\bx}_i} = \mathbf{y}_{k+1-i}\left[ t_{k+1-i,i}+(\range{1}{L}) \right], i=1,\ldots,k.
\end{equation*}
\end{theorem}

We summarize the bandwidth and computation costs of the shift-XOR elimination:
\begin{itemize}
\item First, the algorithm has no bandwidth overhead as the number of input bits is the same as the number of bits to solve. In contrast, the bandwidth overhead of the zigzag algorithm has $\sum_{i=1}^{k}t_{i,k}$ bits.
\item Second, as the algorithm can be implemented in-place, no extra storage space is required to store the intermediate XOR results. The algorithm only needs a small constant number (independent of $k$ and $L$) of intermediate integer variables. (The values $t_{i,j}$, $L_b$ and $L_{1:b}$ are constants that included as a part of the program that implements the algorithm.) Therefore, the shift-XOR elimination uses $O(1)$ auxiliary integer variables. In contrast,  the zigzag
  algorithm needs $O(kL)$ auxiliary integer variables, and $kL$ bits to store the intermediate XOR results.
\item Third, the number of XOR operations used by Algorithm~\ref{alg::mds} is the same as the number of XOR operations used to generate $\hat{\bx}_1, \ldots, \hat\bx_k$ from $\bx_1,\ldots,\bx_k$, and is less than $k(k-1)L$. The number of integer operations used in the algorithm (for calculating back substitution positions) is $O(k^2L)$. Similarly,  the zigzag algorithm needs $k(k+1)L$ XOR operations and $O(k^2L)$ integer operations.
\end{itemize}

To conclude this section, we remark that there are other choices of the subsequences that can guarantee the solvability. But different subsequences may result in different order of the bits to solve, which we would not explore in this paper. For example, if we use $\textbf{y}_u[t_{u,u}+(1:L)]$, $u=1,\ldots,k$, each sequence $\bx_i$ can be solved from the last bit to the first bit.

\section{Decoding and Repair Schemes of Shift-XOR MBR Codes}
\label{sec::mbr}

In this section, we discuss the decoding and repair schemes for the shift-XOR MBR codes described in Section~\ref{sec:constr-xor-mbr}. Our schemes decompose the decoding/repair problem into a sequence of systems of shift-XOR equations, each of which can be solved efficiently using the shift-XOR elimination discussed in the last section.

Let $m, n$ be positive integers, and let $\mathbf{A}=(a_{i,j})$ be an
$m\times n$ matrix. We define some notations to represent submatrices
of $\mathbf{A}$. For two subsets $I\subset \{1,\ldots,n\}$ and
$J\subset \{1,\ldots,m\}$, let $\mathbf{A}_{I}$ (resp. $\mathbf{A}^{J}$) be the submatrix of $\mathbf{A}$ formed by
all the columns (resp. rows) with indices in $I$ (resp. $J$).
Following these notations, $\mathbf{A}^J_I$ is the submatrix of $\mathbf{A}$ formed by the entries on the rows in $J$ and columns in $I$. When
$I=\{i\}$ (resp. $J=\{j\}$), we also write $\mathbf{A}_i$
(resp. $\mathbf{A}^j$) for convenience. These submatrix notations should not be confused with the matrix entries (e.g., $a_{i,j}$), which are specified case by case before using.
%For example, $\mathbf{A}_{i}^{1:j}$ is the submatrix of $\mathbf{A}$ formed by the $i$-th column and the first $j$ rows.

\subsection{Decoding Scheme of Shift-XOR MBR Codes}
\label{sec::mbr_decode}

Consider an $[n,k,d]$ shift-XOR MBR codes.
According to the encoding \eqref{equ::form} with the symmetric matrix $\mathbf{M}=(\mathbf{m}_{i,j})$ in \eqref{eqn:message_matrix}, we get
\begin{equation}
  \label{eq:3}
  \bY = \mathbf{\Psi}
  \begin{bmatrix}
 \mathbf{S} & \mathbf{T}\\
 \mathbf{T}^\top & \mathbf{O}
 \end{bmatrix},
\end{equation}
where $\mathbf{\Psi} = (z^{t_{i,j}})$ is an $n\times d$ matrix satisfying the RID property. We see that for $1\leq i\leq j\leq k$, the $(i,j)$ entry of the symmetric matrix $\mathbf{S}$ is $\mathbf{m}_{i,j}$ and for $1\leq i\leq k, 1\leq j\leq d-k$, the $(i,j)$ entry of $\mathbf{T}$ is $\mathbf{m}_{i,j+k}$. Due to the symmetry of $\mathbf{S}$, there are totally $B = \frac{1}{2}(k+1)k+k(d-k)$ message sequences to decode.
Substituting the entries $\mathbf{m}_{i,j}$ into \eqref{eq:3}, the $(i,j)$ entry of $\bY$ is
\begin{equation}
    \label{equ::mbr_enc}
    \by_{v,u}[l]=\begin{cases} \displaystyle{\sum_{j=1}^{u}\mathbf{m}_{j,u}[l-t_{v,j}]+\sum_{j=u+1}^{d}\mathbf{m}_{u,j}[l-t_{v,j}]}, &  1\leq u\leq k \\
    \displaystyle{\sum_{j=1}^{k}\mathbf{m}_{j,u}[l-t_{v,j}]}, & k<u\leq d
    \end{cases}
\end{equation}
where $1\leq l\leq L+t_{v,d}$.

\subsubsection{Decomposition of Decoding Problem}
\label{sec:overview-our-scheme}

Our scheme decodes the message matrix
$\mathbf{M}$ by first decoding the $k\times (d-k)$ submatrix $\mathbf{T}$ and then
decoding the $k\times k$ symmetric submatrix $\mathbf{S}$.
For each $j=1,\ldots,d-k$, we have the system of shift-XOR equations
\begin{equation}\label{eq:syst}
  \mathbf{Y}_{j+k} = \mathbf{\Psi}_{\range{1}{k}} \mathbf{T}_j,
\end{equation}
where $\mathbf{\Psi}_{\range{1}{k}}$ satisfies the RID property. Using the shift-XOR elimination, $\mathbf{T}_j$ can be decoded from any $k$ rows of \eqref{eq:syst}. The $d-k$ columns of $\mathbf{T}$ can be decoded one-by-one or in parallel.

After decoding $\mathbf{T}$, we continue to decode the $k$ columns
of $\mathbf{S}$ sequentially with the column indices in descending order.
Fix $u$ with $1\leq u \leq k$. We have, according to \eqref{eq:3},
\begin{IEEEeqnarray*}{rCl}
  \mathbf{Y}_u & = & \mathbf{\Psi}_{\range{1}{k}} \mathbf{S}_u +
  \mathbf{\Psi}_{\range{(k+1)}{d}} (\mathbf{T}^u)^\top \\
  & = &
  \mathbf{\Psi}_{\range{1}u} \mathbf{S}_{u}^{\range{1}{u}} + \mathbf{\Psi}_{\range{(u+1)}{k}} \mathbf{S}_{u}^{\range{(u+1)}{k}} +
  \mathbf{\Psi}_{\range{(k+1)}{d}} (\mathbf{T}^u)^\top. \IEEEeqnarraynumspace
\end{IEEEeqnarray*}
Then we have
\begin{equation}
\label{eq:dids}
  \mathbf{Y}_u - \mathbf{\Psi}_{\range{(u+1)}{k}} \mathbf{S}_{u}^{\range{(u+1)}{k}} - \mathbf{\Psi}_{\range{(k+1)}{d}} (\mathbf{T}^u)^\top = \mathbf{\Psi}_{\range{1}u} \mathbf{S}_{u}^{\range{1}{u}}.
\end{equation}
As $\mathbf{\Psi}_{\range{1}u}$ satisfies the RID property, \eqref{eq:dids} can be viewed as a system of shift-XOR equations generated by $\mathbf{\Psi}_{\range{1}u}$ of the input $\mathbf{S}_{u}^{\range{1}{u}}$.
When $u=k$, $\mathbf{S}_{k}$ can be decoded using the shift-XOR elimination on any $k$ rows of $\mathbf{Y}_u- \mathbf{\Psi}_{\range{(k+1)}{d}} (\mathbf{T}^u)^\top$. When $u<k$, suppose that the columns of $\mathbf{S}$ with indices larger than $u$ have all been decoded.
Then $\mathbf{S}_{u}^{\range{1}{u}}$ can be decoded using the shift-XOR elimination on any $u$ rows of \eqref{eq:dids}, where $\mathbf{S}_{u}^{\range{(u+1)}{k}}=\left(\mathbf{S}_{\range{(u+1)}{k}}^u\right)^{\top}$ have been decoded.

\subsubsection{Decoding Scheme}

%The sequences of each row of $\bY$ are stored at one storage node.
Our decoding scheme is able to decode the message sequences by retrieving data from any $k$ out of the $n$ nodes. 
Now we give the details of our decoding scheme, which consists of two stages: the transmission stage and the decoding stage. 
% \subsubsection{Transmission Stage}
% \label{sec:stage1}
In the transmission stage, $k$ storage nodes are chosen. Let the indices of the $k$ nodes be $i_1, i_2, \ldots, i_k$, where $i_1 > i_2
>\cdots > i_k$. For $v \in 1:k$ and $u\in v:d$, define subsequence
\begin{equation}
\label{equ::mbrhat}
  \hat{\mathbf{m}}_{v,u} = \mathbf{y}_{i_v,u}\left[t_{i_v,v}+(\range{1}L)\right],
\end{equation}
which is of $L$ bits.
For $v \in 1:k$, node $i_v$ transmits the subsequences $\hat{\mathbf{m}}_{v,u}$, $u = v,\ldots, d$ to the decoder. 
Substituting \eqref{equ::mbr_enc} into \eqref{equ::mbrhat}, we have for $1\leq l \leq L$,
\begin{equation}
\label{eq:mbrhat2}
 \hat{\mathbf{m}}_{v,u}[l]=\begin{cases}
    \displaystyle{\mf{m}_{v,u}[l]+\sum_{j=1,j\neq v}^{u}\mf{m}_{j,u}[l+t_{i_v,v}-t_{i_v,j}]+\sum_{j=u+1}^{d}\mf{m}_{u,j}[l+t_{i_v,v}-t_{i_v,j}]}, &  1\leq v\leq u\leq k, \\
    \displaystyle{\mf{m}_{v,u}[l]+\sum_{j=1,j\neq v}^{k}\mf{m}_{j,u}[l+t_{i_v,v}-t_{i_v,j}]}, & 1\leq v\leq k,k<u\leq d,
    \end{cases}
\end{equation}
where we see that $\hat{\mf{m}}_{v,u}$ involves  all the bits in $\mf{m}_{v,u}$. 

The decoding stage consists of two steps. In the first step,
matrix $\mathbf{T}$ is decoded using $\hat{\mathbf{m}}_{v,u}$, $1\leq v \leq k, k+1\leq u \leq d$. In the second step, matrix $\mathbf{S}$ is decoded using $\hat{\mathbf{m}}_{v,u}$, $1\leq v \leq u \leq k$. A pseudocode of the decoding scheme is shown in
Algorithm~\ref{alg::mbrDecoding}, which also demonstrates an in-place implementation of the decoding algorithm.
\begin{itemize}
\item[Step 1:] This step contains $d- k$ iterations. For each $u\in \range{(k+1)}{d}$, performing the shift-XOR elimination on
$\hat{\mathbf{m}}_{1,u},\hat{\mathbf{m}}_{2,u},\ldots,\hat{\mathbf{m}}_{k,u}$, 
we can decode
$\mathbf{m}_{1,u},\mathbf{m}_{2,u},\ldots,\mathbf{m}_{k,u}$ (ref. Line~2 in Algorithm~\ref{alg::mbrDecoding}). For in-place implementation, $\mathbf{m}_{v,u}$ is stored at the same storage place of storing $\hat{\mathbf{m}}_{v,u}$.
For any $v$ with $1\leq v\leq k$, $\mathbf{m}_{v,u}$ is involved in
generating $\hat{\mf{m}}_{w,v}$, $1\leq w\leq v$ as in~\eqref{eq:mbrhat2}.
So the value of $\mathbf{m}_{v,u}$ is substituted in
$\hat{\mathbf{m}}_{w,v}$ (ref. Line~3--5 in Algorithm~\ref{alg::mbrDecoding}). After the
substitution, ${{\hat{\mathbf{m}}}_{w,v}}$ with $1\leq w\leq v\leq k$ is  only related to
${{\mathbf{{m}}}_{v',u'}}, 1\leq v'\leq u'\leq k$.

\item[Step 2:] This step contains $k-1$
iterations. For iteration $u$ from $k$ down to $2$, performing the shift-XOR elimination of $\hat{\mathbf{m}}_{1,u},\hat{\mathbf{m}}_{2,u},\ldots,\hat{\mathbf{m}}_{u,u}$
to decode $\mathbf{m}_{1,u},\mathbf{m}_{2,u},\ldots ,\mathbf{m}_{u,u}$ (ref. Line~7 in Algorithm~\ref{alg::mbrDecoding}).
Since $\mathbf{m}_{v,u}$ $(1\le v\leq u-1)$ is involved in generating
the bits in $\hat{\mf{m}}_{w,v}$ $(1\le w\leq v)$, the sequence
$\mathbf{m}_{v,u}$ $(1\le v\leq u-1)$ is then substituted into $\hat{\mathbf{m}}_{w,v}$ $(1\le
w\leq v)$, according to~\eqref{eq:mbrhat2} 
(ref. Line~8--10 in Algorithm~\ref{alg::mbrDecoding}). After the substitution, ${{\hat{\mathbf{m}}}_{w,v}}$
($1\leq w\leq v\leq u-1$) is only related to
${\mathbf{{m}}_{v',u'}}, 1\leq v'\leq u' \leq u-1$.
\end{itemize}
After Step 2, $\mathbf{m}_{v,u}$ ($1\leq v\leq k, v\leq u\leq d$) is decoded and stored at the same storage place as $\hat{\mathbf{m}}_{v,u}$.

% Similar as Algorithm~\ref{alg::mds}, Algorithm~\ref{alg::mbrDecoding} also has an  in-place implementation, so that the storage space of $\hat{\mf{m}}_{u,v}[l]$ becomes ${\mf{m}_{u,v}}[l]$ after decoding.

\begin{algorithm}[t]
\caption{Decoding algorithm for shift-XOR MBR codes with in-place implementation. After the execution, the output value $\mathbf{m}_{v,u}[l]$ is stored at the same storage space as $\hat{\mathbf{m}}_{v,u}[l]$.}
\label{alg::mbrDecoding}
\begin{algorithmic}[1]
\REQUIRE coded sequences ${{\hat{\mathbf{m}}}_{v,u}}$, $1\leq v\leq k, v\leq u\leq d$, and corresponding node indices ${{i}_{j}}$ ($1\le j\le k$), which satisfies ${{i}_{1}}>{{i}_{2}}>\cdots >{{i}_{k}}$.
\ENSURE decoded message sequences ${{{\mathbf{m}}}_{v,u}}$, $1\leq v\leq k, v\leq u\leq d$.

(\textbf{Step 1:}  Decode $\mathbf{T}$)
\FOR {$u\leftarrow d$ down to $k+1$}
\STATE Decode $({{\mathbf{m}}}_{v,u}, 1\leq v\leq k)$ by executing  Algorithm~\ref{alg::mds} on  $\hat{\mathbf{m}}_{1,u},\hat{\mathbf{m}}_{2,u},\ldots,\hat{\mathbf{m}}_{k,u}$ with  indices $i_1, i_2, \ldots, i_k$.
    \FOR {$v\leftarrow 1:k$}
        \FOR {$w\leftarrow 1:v$}
\STATE        $\hat{\mathbf{m}}_{w,v}\left[\range{(t_{i_w,u}-t_{i_w,w}+1)}{L}\right] \oplus \leftarrow  {\mathbf{m}}_{v,u} \left[\range{1}{(L-t_{i_w,u}+t_{i_w,w})}\right].$ (in-place back substitution)
        \ENDFOR
    \ENDFOR
\ENDFOR

(\textbf{Step 2:} Decode $\mathbf{S}$)
\FOR { $u\leftarrow k$ down to $2$ }
\STATE Decode $({\mathbf{m}}_{1,u},{\mathbf{m}}_{2,u},\ldots,{\mathbf{m}}_{u,u})$ executing Algorithm~\ref{alg::mds} on  $\hat{\mathbf{m}}_{1,u},\hat{\mathbf{m}}_{2,u},\ldots,\hat{\mathbf{m}}_{u,u}$ with indices $i_1, i_2, \ldots, i_u$.
    \FOR {$v\leftarrow 1:(u-1)$}
        \FOR {$w\leftarrow 1:v$}
\STATE        $\hat{\mathbf{m}}_{w,v}\left[\range{(t_{i_w,u}-t_{i_w,w}+1)}L\right] {\oplus} \leftarrow  {\mathbf{m}}_{v,u} \left[\range{1}{(L-t_{i_w,u}+t_{i_w,w})}\right].$ (in-place back substitution)
        \ENDFOR
    \ENDFOR
\ENDFOR
\end{algorithmic}
\end{algorithm}

\begin{example}
Consider the $\left[ 6,3,4 \right]$ shift-XOR MBR code in Example~\ref{exam::mbr}.
Suppose the transmission stage chooses nodes $1$, $3$ and $4$, i.e., $i_1 = 4$, $i_2 = 3$, and $i_3 = 1$. Node $i_1=4$ transmits $\hat{\mathbf{m}}_{1,i} = \mathbf{y}_{4,i}[1:L]$, $i=1,\ldots,4$ to the decoder. Node $i_2=3$ transmits $\hat{\mathbf{m}}_{2,i}=\mathbf{y}_{3,i}[3:(L+2)]$, $i=2,3,4$ to the decoder. Node $i_3=1$ transmits $\hat{\mathbf{m}}_{3,i}=\mathbf{y}_{1,i}[1:L]$, $i=3,4$ to the decoder.

  Since $d=k+1$, Step 1 of Algorithm~\ref{alg::mbrDecoding} has only one iteration. According to~\eqref{eq:mbrhat2}, we have for $1\leq l\leq L$ 
  \begin{eqnarray*}
      \hat{\mathbf{m}}_{1,4}[l]&=&\mathbf{m}_{1,4}[l]+\mathbf{m}_{2,4}[l-3]+\mathbf{m}_{3,4}[l-3],\\
      \hat{\mathbf{m}}_{2,4}[l]&=&\mathbf{m}_{2,4}[l]+\mathbf{m}_{1,4}[l+2]+\mathbf{m}_{3,4}[l-2],\\
      \hat{\mathbf{m}}_{3,4}[l]&=&\mathbf{m}_{3,4}[l]+\mathbf{m}_{1,4}[l]+\mathbf{m}_{2,4}[l].
  \end{eqnarray*}
% which are shift-XOR equations of $\mathbf{m}_{1,4}$, $\mathbf{m}_{2,4}$ and $\mathbf{m}_{3,4}$.
Performing the shift-XOR elimination on  $\hat{\mathbf{m}}_{1,4}$, $\hat{\mathbf{m}}_{2,4}$ and $\hat{\mathbf{m}}_{3,4}$ we obtain $\mathbf{m}_{1,4}$, $\mathbf{m}_{2,4}$ and $\mathbf{m}_{3,4}$. For $1\leq l\leq L$,
  \begin{eqnarray*}
  \hat{\mathbf{m}}_{1,1}[l]&=&\mathbf{m}_{1,1}[l]+\mathbf{m}_{1,2}[l-3]+\mathbf{m}_{1,3}[l-6]+\mathbf{m}_{1,4}[l-9],\\
  \hat{\mathbf{m}}_{1,2}[l]&=&\mathbf{m}_{1,2}[l]+\mathbf{m}_{2,2}[l-3]+\mathbf{m}_{2,3}[l-6]+\mathbf{m}_{2,4}[l-9],\\
  \hat{\mathbf{m}}_{1,3}[l]&=&\mathbf{m}_{1,3}[l]+\mathbf{m}_{2,3}[l-3]+\mathbf{m}_{3,3}[l-6]+\mathbf{m}_{3,4}[l-9],\\
  \hat{\mathbf{m}}_{2,2}[l]&=&\mathbf{m}_{2,2}[l]+\mathbf{m}_{1,2}[l+2]+\mathbf{m}_{2,3}[l-2]+\mathbf{m}_{2,4}[l-4],\\
  \hat{\mathbf{m}}_{2,3}[l]&=&\mathbf{m}_{2,3}[l]+\mathbf{m}_{1,3}[l+2]+\mathbf{m}_{3,3}[l-2]+\mathbf{m}_{3,4}[l-4],\\
  \hat{\mathbf{m}}_{3,3}[l]&=&\mathbf{m}_{3,3}[l]+\mathbf{m}_{1,3}[l]+\mathbf{m}_{2,3}[l]+\mathbf{m}_{3,4}[l].
\end{eqnarray*}
The decoder further substitutes
  $\mathbf{m}_{1,4}$ into $\hat{\mathbf{m}}_{1,1}$, substitutes $\mathbf{m}_{2,4}$ into $\hat{\mathbf{m}}_{1,2}$ and $\hat{\mathbf{m}}_{2,2}$, and substitutes
  $\mathbf{m}_{3,4}$ into $\hat{\mathbf{m}}_{1,3}$,
  $\hat{\mathbf{m}}_{2,3}$ and $\hat{\mathbf{m}}_{3,3}$ correspondingly. We use the same notation to denote the sequences after substitution.

  Step~2 of Algorithm~\ref{alg::mbrDecoding} has two iterations with
  $u=3$ and $u=2$ respectively. In the $u=3$ iteration, we have for $1\leq l\leq L$ 
  \begin{eqnarray*}
      \hat{\mathbf{m}}_{1,3}[l]&=&\mathbf{m}_{1,3}[l]+\mathbf{m}_{2,3}[l-3]+\mathbf{m}_{3,3}[l-6],\\
      \hat{\mathbf{m}}_{2,3}[l]&=&\mathbf{m}_{2,3}[l]+\mathbf{m}_{1,3}[l+2]+\mathbf{m}_{3,3}[l-2],\\
      \hat{\mathbf{m}}_{3,3}[l]&=&\mathbf{m}_{3,3}[l]+\mathbf{m}_{1,3}[l]+\mathbf{m}_{2,3}[l],
  \end{eqnarray*}
  which can be solved by shift-XOR elimination to obtain 
  % Algorithm~\ref{alg::mds} converts $\hat{\mathbf{m}}_{1,3}$,
  % $\hat{\mathbf{m}}_{2,3}$ and $\hat{\mathbf{m}}_{3,3}$ to
  $\mathbf{m}_{1,3}$, $\mathbf{m}_{2,3}$ and $\mathbf{m}_{3,3}$. Similarly, the decoder substitutes $\mathbf{m}_{1,3}$ into $\hat{\mathbf{m}}_{1,1}$, substitutes $\mathbf{m}_{2,3}$ into
  $\hat{\mathbf{m}}_{1,2}$ and $\hat{\mathbf{m}}_{2,2}$.
  In the $u=2$ iteration, the shift-XOR elimination is performed on $\hat{\mathbf{m}}_{1,2}$ and $\hat{\mathbf{m}}_{2,2}$ to decode $\mathbf{m}_{1,2}$ and $\mathbf{m}_{2,2}$. The decoder then substitutes $\mathbf{m}_{1,2}$ into $\hat{\mathbf{m}}_{1,1}$, which becomes ${\mathbf{m}}_{1,1}$.
\end{example}

\subsubsection{Complexity Analysis}
\textbf{Time Complexity:} In Step~1, decoding each column of
$\mathbf{T}$ costs $k(k-1)L$ XOR operations by the shift-XOR
elimination (Line 2~in Algorithm~\ref{alg::mbrDecoding}).  Noting that
in Line~5 of Algorithm~\ref{alg::mbrDecoding}, the sequences for back
substitution are shorter than $L$, and hence substituting each column
of $\mathbf{T}$ into other retrieved sequences takes less than
$\frac{1}{2}k(k+1)L$ XOR operations (Line 3 -- 5 in
Algorithm~\ref{alg::mbrDecoding}). There are $\left(d-k\right)$
iterations in this step, so the number of XOR operations $T_1$ required in
Step~1 satisfies
\begin{eqnarray*}
T_{1} \!\! &\!\! < \!\!&\!\! (d-k)\left(k(k-1)+\frac{1}{2}k(k+1)\right)L \\
&\!\! = \!\!&\!\! \frac{1}{2}\left(d-k\right)\left(3k-1\right)kL.
\end{eqnarray*}
In Step~2, similarly, an iteration $u \in \range{2}{k}$ needs $u(u-1)L$ XOR
operations to execute Algorithm~\ref{alg::mds} and less than 
$\frac{1}{2}u\left(u-1\right)L$ XOR operations to substitute the
decoded sequences. Therefore, the number of XOR operations $T_2$ in the second step satisfies
\begin{eqnarray*}
T_{2} \!\! &\!\! < \!\!&\!\! \sum_{u=2}^{k}\left(u(u-1)L+\frac{1}{2}u(u-1)L\right) \\
&\!\! = \!\!&\!\! \frac{1}{2}(k-1)k(k+1)L. \nonumber
\end{eqnarray*}
Therefore, the total number of XOR operations is
\begin{eqnarray*}
T_1+T_2 \!\! &\!\! < \!\!&\!\! \left(\left(\frac{3}{2}d-k\right)k-\frac{1}{2}(d-k+1)\right)kL.
\end{eqnarray*}
So the time complexity is $O(dk^2L)$.

\textbf{Space Complexity:} Same as the shift-XOR elimination, Algorithm~\ref{alg::mbrDecoding} can be implemented in-place, so that the output sequences take the same storage space as the input sequences. No extra space is required to store the intermediate XOR results. Only $O(1)$ integer auxiliary variables are required by Algorithm~\ref{alg::mbrDecoding}. 

\textbf{Bandwidth Overhead:} Our algorithm consumes exactly $BL$ bits from the storage nodes to decode the $BL$ bits of the message sequences. Therefore, our decoding algorithm has zero bandwidth overhead.

\subsection{Repair Scheme for Shift-XOR MBR Codes}
\label{sec::mbr_repair}

This section introduces the repair scheme for the  $[n,k,d]$ shift-XOR MBR code described in Section~\ref{sec:constr-xor-mbr}.
Suppose that node $i$ fails. Our repair scheme generates a new storage
node that stores $d$ sequences 
\begin{equation}
    \label{eq:mbr_repair}
    \mathbf{Y}^i = \mathbf{\Psi}^{i}\mathbf{M}=[\mathbf{y}_{i,1},\ldots,\mathbf{y}_{i,d}],
\end{equation}
same as the $d$ sequences stored at node $i$. Note that a sequence in $\mathbf{Y}^i$ has $L+t_{i,d}$ bits.
Recall that each storage node $j\in 1:n$ stores the sequences $\mathbf{\Psi}^{j}\mf M$, and hence can compute locally the sequence
\begin{equation}\label{eq:r}
\mathbf{r}_j=\mathbf{\Psi}^{j}\mathbf{M}\left(\mathbf{\Psi}^i\right)^{\top}=\mathbf{\Psi}^{j}(\mathbf{\Psi}^i\mathbf{M})^{\top} = \mathbf{\Psi}^{j} (\mf Y^i)^\top,
\end{equation}
which is a shift-XOR equation of $\mathbf{Y}^i$. Therefore, using the shift-XOR elimination, $\mathbf{Y}^i$ can be decoded from $\mathbf{r}_j$ of any $d$ nodes $j$.

% The sequences stored at node $j\in[1,n]$ are $\mathbf{\Psi}^{j}M$. Then the inner product can be computed at node $j$. As $\mathbf{r}_j$ can be viewed as a shift-XOR linear system of $\mathbf{\Psi}_i\mathbf{M}$, which consists of $d$ sequences, any selected $d$ $r_{j}$ can be used to repair $\mathbf{\Psi}_i\mathbf{M}$ by applying Algorithm~\ref{alg::mds}.

Specifically, the repair scheme includes two stages: the transmission stage and the decoding stage.
In the transmission stage, $d$ \emph{helper nodes} are chosen to repair node $i$, which have the indices 
 $i_1, i_2, \ldots, i_d$ where $i_{1}>i_{2}>\cdots >i_{d}$.
Each node $i_{v}$  ($1 \le v \le d$) transmits a subsequence of $\mathbf{r}_{i_v}$ (defined in \eqref{eq:r}) 
\begin{equation}
\label{eq:mhat_mbr_repair}
\hat{\mathbf{m}}_v =
\mathbf{r}_{i_v}[\range{(1+t_{i_v,v})}{(L+t_{i,d}+t_{i_v,v})}]
\end{equation}
to the new node $i$ for repairing. Note that $\hat{\mathbf{m}}_v$ has exactly the same number of bits as the sequences to repair.
In the decoding stage, the new node $i$ has received sequences $\hat{\mathbf{m}}_v$, $v=1,\ldots,d$, and performs the shift-XOR elimination to decode $\mathbf{Y}^i$.

% Substitute~\eqref{eq:mbr_repair} into~\eqref{eq:mhat_mbr_repair}, we have for $1\leq l\leq L+t_{i,d}$
% \begin{equation*}
%     \hat{\mathbf{m}}_{v}[l]=\mathbf{y}_{i,v}[l]+\sum_{j\neq v}\mathbf{y}_{i,j}[l-t_{i_v,j}+t_{i_v,v}].
% \end{equation*}
% In decoding stage, Algorithm~\ref{alg::mds} is executed to convert $\hat{\mathbf{m}}_v$ into $\mathbf{y}_{i,v}$ in-place.

\begin{example}
 Consider again $\left[6,3,4 \right]$ MBR code in Example~\ref{exam::mbr}, where the message matrix and generator matrix can be found.
 This example will show the transmission stage and decoding stage to repair node $3$ by connecting helper nodes $1$, $2$, $4$ and $5$, i.e., $i_{1}=5$, $i_{2}=4$, $i_{3}=2$, $i_{4}=1$. So 
\begin{eqnarray*}
%\mathbf{r}_{i_{1}}=
\mathbf{r}_{{5}}\!\!&\!\! = \!\!&\!\! \mathbf{y}_{3,1}+z^{4}\mathbf{y}_{3,2}+z^{8}\mathbf{y}_{3,3}+z^{12}\mathbf{y}_{3,4},\\
%\mathbf{r}_{i_{2}}=
\mathbf{r}_{{4}}\!\!&\!\! = \!\!&\!\! \mathbf{y}_{3,1}+z^{3}\mathbf{y}_{3,2}+z^{6}\mathbf{y}_{3,3}+z^{9}\mathbf{y}_{3,4}, \\
%\mathbf{r}_{i_{3}}=
\mathbf{r}_{{2}}\!\!&\!\! = \!\!&\!\! \mathbf{y}_{3,1}+z^{1}\mathbf{y}_{3,2}+z^{2}\mathbf{y}_{3,3}+z^{3}\mathbf{y}_{3,4},\\
%\mathbf{r}_{i_{4}}=
\mathbf{r}_{{1}}\!\!&\!\! = \!\!&\!\! \mathbf{y}_{3,1}+\mathbf{y}_{3,2}+\mathbf{y}_{3,3}+\mathbf{y}_{3,4}.
\end{eqnarray*}
Then the sequences transmitted to the new node $3$ are
  \begin{IEEEeqnarray*}{rCl}
  \hat{\mathbf{m}}_{1} &=& \mathbf{r}_{5}[\range{1}{(L+6)}] \\
%  & = & (\mathbf{y}_{3,1}+z^{4}\mathbf{y}_{3,2}+z^{8}\mathbf{y}_{3,3}+z^{12}\mathbf{y}_{3,4})[\range{1}{L+6}], \\
  \hat{\mathbf{m}}_{2} &=& \mathbf{r}_{4}[\range{4}{(L+9)}] \\
%  & = & (z^{-3}\mathbf{y}_{3,1}+\mathbf{y}_{3,2}+z^{3}\mathbf{y}_{3,3}+z^{6}\mathbf{y}_{3,4})[\range{1}{L+6}], \\
  \hat{\mathbf{m}}_{3} &=& \mathbf{r}_{2}[\range{3}{(L+8)}] \\
%  & = & (z^{-2}\mathbf{y}_{3,1}+z^{-1}\mathbf{y}_{3,2}+\mathbf{y}_{3,3}+z^{1}\mathbf{y}_{3,4})[\range{1}{L+6}], \\
  \hat{\mathbf{m}}_{4} &=& \mathbf{r}_{1}.
%  & = & \mathbf{y}_{3,1}+\mathbf{y}_{3,2}+\mathbf{y}_{3,3}+\mathbf{y}_{3,4}.
\end{IEEEeqnarray*}
Applying Algorithm~\ref{alg::mds} on
$\hat{\mathbf{m}}_{1}$, $\hat{\mathbf{m}}_{2}$, $\hat{\mathbf{m}}_{3}$ and $\hat{\mathbf{m}}_{4}$, we repair
$\mathbf{y}_{3,1}$, $\mathbf{y}_{3,2}$, $\mathbf{y}_{3,3}$ and $\mathbf{y}_{3,4}$ at the new
node $3$.
\end{example}

\textbf{Time Complexity:}
The repair computation cost involves two parts: The first part is the computation at the $d$ helper nodes, and the second part is the computation at the repaired node.
At each helper node, $(d-1)\left(L+t_{i,d}\right)$ XOR operations are used. So the total number of XOR operations at all the helper nodes is $d\left(d-1\right)(L+t_{i,d})$.
 The number of XOR operations of the second part is $d(d-1)\left(L+t_{i,d}\right)$,
 according to the analysis of Algorithm~\ref{alg::mds}.
 Totally, the number of XOR operations of MBR codes repair is $2d(d-1)(L+t_{i,d}) = 2d(d-1)L + O(nd^3)$.

\textbf{Space Complexity:}
 In the repaired node, one shift-XOR elimination is performed and $O(1)$ auxiliary integer variables are required.

\textbf{Bandwidth:}
Our algorithm consumes exactly $L+t_{i,d}$ bits from $d$ helper nodes to repair the $d(L+t_{i,d})$ bits of node $i$, and hence has zero bandwidth overhead.

\section{Decoding and Repair Schemes of Shift-XOR MSR Codes}
\label{sec::msr}

In this section, we discuss the decoding and repair schemes for the shift-XOR MSR codes described in Section~\ref{sec::cons-xor-msr}. Similar to those of the shift-XOR MBR codes, our schemes decompose the decoding/repair problem into a sequence of systems of shift-XOR equations, each of which can be solved efficiently using the shift-XOR elimination.

\subsection{Decoding Scheme of Shift-XOR MSR Codes}
\label{sec::dec-msr}

Consider  an $\left[ n,k,d \right]$ shift-XOR MSR code, where $d=2k-2$ and $\alpha=k-1$. Substituting the message matrix $\mf M$ in \eqref{eq:msr-m} and the generator matrix $\mf \Psi$ in \eqref{eq:msr:p} into the general encoding formula \eqref{equ::form}, we obtain
\begin{equation}\label{eq:dmsr}
  \mf Y^{i} = \mf\Phi^{i}\mf S + z^{\lambda_i}\mf\Phi^{i}\mf T,
\end{equation}
% \begin{equation*}
%   \mf Y = \mf\Phi\mf S + \mf\Lambda\mf\Phi\mf T.
% \end{equation*}
which is the $i$-th row of $\mf Y$ stored at storage node $i$.
Denote the $(v,u)$ entry of $\mf S$ and $\mf T$ as $\mf s_{v,u}$ and $\mf t_{v,u}$, respectively. Due to symmetry, the decoding problem is to solve the message sequences $\mf s_{v,u}$, $\mf t_{v,u}$, $1\leq v \leq u \leq \alpha$, totally $B = \alpha(\alpha+1) = (k-1)k$ sequences. % using the sequences stored at any $k$ out of the $n$ storage nodes. 

\subsubsection{Decomposition of Decoding Problem}

Suppose the indices of the $k$ nodes chosen for decoding are $i_1,i_2,\ldots,i_k$, where $i_{1}>i_{2}>\cdots >i_{k}$, so that the decoder can retrieve $\mf Y^{i_{u}}$, $u=1,\ldots, k$. Denote for $1\leq u \neq v \leq k$,
\begin{IEEEeqnarray}{rCl}
  \mf c_{u,v} & = & \mf Y^{i_u} (\mathbf{\Phi}^{i_{v}})^\top, \label{eq:c} \\
  \mf p_{u,v} & = & \mf\Phi^{i_u} \mf S (\mathbf{\Phi}^{i_{v}})^\top, \label{eq:p}\\
  \mf q_{u,v} & = & \mf\Phi^{i_u} \mf T (\mathbf{\Phi}^{i_{v}})^\top. \label{eq:q}
\end{IEEEeqnarray}
Here $\mf c_{u,v}$ can be computed by the decoder using the sequences it retrieved.
Due to the symmetry of $\mf S$ and $\mf T$, we have $\mf p_{u,v} = \mf p_{v,u}$ and $\mf q_{u,v}=\mf p_{v,u}$.
Briefly, the decoding problem is decomposed into the following two steps:
\begin{itemize}
\item First, $\mf p_{u,v}$, $\mf q_{u,v}$, $1\leq u < v\leq k$ are solved using $\mf c_{u,v}$, $1\leq u \neq v \leq k$.
\item Second, $\mf S$ is solved using $\mf p_{u,v}$, $1\leq u < v\leq k$, and $\mf T$ is solved using $\mf q_{u,v}$, $1\leq u < v\leq k$.
\end{itemize}
Let us elaborate these two steps.

\textbf{Step 1:}
For $1\leq v<u\leq k$, we have the shift-XOR equations (obtained by \eqref{eq:dmsr} -- \eqref{eq:q})
\begin{equation}
\label{equ::generator_msr_1}
\begin{bmatrix}
\mathbf{c}_{u,v} \\
\mathbf{c}_{v,u}
\end{bmatrix} =\begin{bmatrix}
\mathbf{p}_{u,v}+z^{\lambda_{i_u}}\mathbf{q}_{u,v}\\
\mathbf{p}_{v,u}+z^{\lambda_{i_v}}\mathbf{q}_{v,u}\\
\end{bmatrix}=
\begin{bmatrix}
\mathbf{p}_{u,v}+z^{\lambda_{i_u}}\mathbf{q}_{u,v}\\
\mathbf{p}_{u,v}+z^{\lambda_{i_v}}\mathbf{q}_{u,v}\\
\end{bmatrix}=
\begin{bmatrix}
{1} & {z^{\lambda_{i_u}}} \\
{1} & {z^{\lambda_{i_{v}}}}
\end{bmatrix}
\begin{bmatrix}
\mathbf{p}_{u,v} \\
\mathbf{q}_{u,v}
\end{bmatrix}.
\end{equation}
Due to the RID property of $\left[\mf\Phi\ \mf\Lambda\mf\Phi\right]$, we have $\lambda_{i_u}t_{i_u,1}-t_{i_u,a} < \lambda_{i_v} t_{i_v,1}-t_{i_v,a}$ and
$t_{i_u,a}-t_{i_u,1} < t_{i_v,a}-t_{i_v,1}$, which implies
$\lambda_{i_u}\neq \lambda_{i_v}$.
Hence, the shift-XOR elimination can be performed on $\mf c_{u,v}$ and $\mf c_{v,u}$ to solve $\mf p_{u,v}$ and $\mf q_{u,v}$, which are of $L+t_{i_u,\alpha}+t_{i_{v},\alpha}$ bits.
According to the discussion in Section~\ref{sec::mds}, only the subsequences
\begin{IEEEeqnarray}{rCl}
\hat{\mathbf{c}}_{u,v} & := & \mathbf{c}_{u,v}[\range{1}{(L+t_{i_u,\alpha}+t_{i_{v},\alpha})}] \label{eq:cuv1} \\ \hat{\mathbf{c}}_{v,u} & := & \mathbf{c}_{v,u}[\lambda_{i_{v}}+(\range{1}{(L+t_{i_u,\alpha}+t_{i_{v},\alpha})})]  \label{eq:cuv2}
\end{IEEEeqnarray}
are needed for the shift-XOR elimination.

\textbf{Step 2:}
Define an $\alpha\times \alpha$ matrix $\tilde{\mathbf{S}} = (\tilde{\mf s}_{v,u})$ as 
\begin{equation}\label{eq:tildes}
  \tilde{\mathbf{S}} = %\mathbf{\Phi}^{i_{v}}
  \begin{bmatrix}
    \mathbf{\Phi}^{i_{1}} \\ \vdots \\ \mathbf{\Phi}^{i_{\alpha}}
  \end{bmatrix}
  \mf S.
\end{equation}
Due to the symmetry of $\mf S$, $(\tilde{\mathbf{S}}^v)^\top = \mf S (\mf\Phi^{i_v})^\top$.
%Recall that $k=\alpha+1$.
For each $v\in 1:\alpha$, form the $\alpha\times \alpha$ system of shift-XOR equations (by \eqref{eq:p} and \eqref{eq:tildes})
\begin{equation}
\label{equ::generator_msr_2}
\begin{bmatrix}
\mathbf{p}_{1,v} \\
\vdots \\
\mathbf{p}_{v-1,v} \\
\mathbf{p}_{v+1,v} \\
\vdots \\
\mathbf{p}_{k,v}
\end{bmatrix}  =
\begin{bmatrix}
\mathbf{\Phi}^{i_{1}} \\
\vdots \\
\mathbf{\Phi}^{i_{v-1}} \\
\mathbf{\Phi}^{i_{v+1}} \\
\vdots \\
\mathbf{\Phi}^{i_{k}} \\
\end{bmatrix}
(\tilde{\mathbf{S}}^{v})^{\top},
\end{equation}
so that $(\tilde{\mathbf{S}}^{v})^{\top}$ can be solved by performing the shift-XOR elimination on the LHS of \eqref{equ::generator_msr_2}. By \eqref{eq:tildes}, we see the maximum length of sequences in $\tilde{\mathbf{S}}^{v}$ is $L'_v = L +  t_{i_{v},\alpha}$. So the shift-XOR elimination only needs the subsequences
\begin{equation*}
  \hat{\mf p}_{u,v} :=
  \begin{cases}
    \mathbf{p}_{u,v}[t_{i_u,u}+(1:L'_v)], & u=1,\ldots, v-1 \\
    \mathbf{p}_{u,v}[t_{i_{u},u-1}+(1:L'_v)], & u=v+1,\ldots, k.
  \end{cases}
\end{equation*}
 % where
  % \begin{equation*}
  % L'_v = L +  t_{i_{v},\alpha}.
  %       \end{equation*}
After solving $\tilde{\mathbf{S}}$, we further solve the system of shift-XOR equations \eqref{eq:tildes},
% \begin{equation}
% \label{equ::generator_msr_3}
% \tilde{\mathbf{S}}^{v} = \mathbf{\Phi}^{i_{v}} \mathbf{S}, \quad v=1,\ldots,\alpha,
% \end{equation}
so that $\mf S_u$ can be decoded by the shift-XOR elimination on
\begin{equation*}
\hat{\mf s}_{v,u} := \tilde{\mathbf{s}}_{v,u}[t_{i_{v},v}+(1:L)], \quad v=1,\ldots, \alpha.  
\end{equation*}

The procedure for solving $\mf T$ is the same and hence is omitted.

\subsubsection{Decoding Scheme}
\label{sec::msr_decode}

Our decoding scheme is able to decode the message sequences by
retrieving data from any $k$ out of the $n$ nodes.  Now we give the
details of our decoding scheme, which consists of two stages: the
transmission stage and the decoding stage. In the transmission stage,
$k$ storage nodes are chosen. Let the indices of the $k$ nodes be
$i_1, i_2, \ldots, i_k$, where $i_1 > i_2 >\cdots > i_k$. 
For $u \in 1:k$, node $i_u$ transmits  $\bY^{i_u}$ to the decoder. 

The decoding stage includes the two steps described above, with the pseudocode in Algorithm~\ref{alg::msrdec}. The algorithm inputs $\bY^{i_u}$, $u=1,\ldots, k$ (totally $B$ sequences) and outputs the $B$ message sequences. But different from the decoding of shift-XOR MBR codes, this algorithm needs extra storage space for the intermediate XOR results.

In Step 1, the algorithm first calculates $\hat{\mf c}_{u,v}$ and $\hat{\mf c}_{v,u}$, $1\leq u<v \leq k$. From \eqref{eq:cuv1} and \eqref{eq:cuv2}, we see the length of $\hat{\mf c}_{u,v}$ and $\hat{\mf c}_{v,u}$ is $L+O(nd)$.
% can be longer than sequences of $\bY^{i_u}$, and hence extra $O(Bnd)$ storage space beyond those for $\bY^{i_u}$ is required
To storage $\hat{\mf c}_{u,v}$ and $\hat{\mf c}_{v,u}$, $1\leq u<v \leq k$, 
% There are $\alpha(\alpha+1)$ sequences, and hence
a space of  $\alpha(\alpha+1)(L+O(nd))$ bits is needed.
In Line~\ref{a3_2}, the shift-XOR elimination is applied on $\hat{\mf c}_{u,v}$ and $\hat{\mf c}_{v,u}$ % (which are subsequences of $\mf c_{u,v}$ and $\mf c_{v,u}$, respectively)
to solve $\mf p_{u,v}$ and $\mf q_{u,v}$ for $1\leq v < u \leq k$. As the shift-XOR elimination is in-place, $\mf p_{u,v}$ and $\mf q_{u,v}$ can take exactly the same storage space as $\hat{\mf c}_{u,v}$ and $\hat{\mf c}_{v,u}$ for $1\leq v < u \leq k$, respectively. Then another space of $\alpha(\alpha-1)$ sequences is needed to store $\mathbf{p}_{u,v}$ and $\mathbf{q}_{u,v}$, $1 \leq u < v \leq \alpha$. %^$u=3,\ldots,k$, $v=2,\ldots,u-1$.
Therefore, after Line \ref{a3_7}, totally $2\alpha^2(L+O(nd))$ bits space is needed.

In Step 2, from Line~\ref{a3_8} to \ref{a3_9}, the shift-XOR elimination is applied on $\hat{\mathbf{p}}_{u,v}$, $u=1,\ldots, v-1,v+1,\ldots,k$ to solve $\tilde{\mf S}^v$ for $v=1,\ldots, \alpha$.
From Line~\ref{a3_10} to \ref{a3_11}, the shift-XOR elimination is applied on $\hat{\mathbf{s}}_{u,v}$, $v=1, 2,\ldots, \alpha$ to solve $\mf S_u$ for $u=1,\ldots, \alpha$. Due to the in-place property of the shift-XOR elimination, no additional space is needed.

Last, $\mf T$ is solved by repeating the above process on $\hat{\mathbf{q}}_{u,v}$ instead of $\hat{\mathbf{p}}_{u,v}$.

\begin{algorithm}[t]
\caption{Decoding algorithm for shift-XOR MSR codes}
\label{alg::msrdec}
\begin{algorithmic}[1]
\REQUIRE coded sequences $\mf Y^{i_u}$, $u=1,\ldots, k$.

\ENSURE  message sequences $\mathbf{s}_{u,v}$ and $\mathbf{t}_{u,v}$, $1\leq u \leq \alpha, 1 \le v\leq \alpha$.

(\textbf{Step~1:} Solve $\mathbf{p}_{v,u}$ and $\mathbf{q}_{v,u}$)

\STATE {Calculate $\hat{\mf c}_{u,v}$ and $\hat{\mf c}_{v,u}$ for $1\leq u<v \leq k$.}

\FOR {$u\leftarrow 2:k$}
    \FOR {$v\leftarrow 1:u-1$}
    \STATE {Apply Algorithm~\ref{alg::mds} on $\hat{\mf c}_{u,v}$ and $\hat{\mf c}_{v,u}$ to solve $\mf p_{u,v}$ and $\mf q_{u,v}$.} \label{a3_2}
    \IF {$u<k$}
    \STATE {${\mathbf{p}}_{v,u}\leftarrow  {\mathbf{p}}_{u,v}$.}
    \STATE {${\mathbf{q}}_{v,u}\leftarrow  {\mathbf{q}}_{u,v}$.} \label{a3_7}
    \ENDIF
    \ENDFOR
\ENDFOR

(\textbf{Step~2:} Decode $\mathbf{S}$ and $\mathbf{T}$)

    \FOR{$v\leftarrow 1:\alpha$} \label{a3_8}
        \STATE {Apply Algorithm~\ref{alg::mds} on
          $\hat{\mathbf{p}}_{u,v}$, $u=1,\ldots, v-1,v+1,\ldots,k$ to solve $\tilde{\mf S}^v$.}\label{a3_9}
    \ENDFOR
    \FOR{$u\leftarrow 1:\alpha$} \label{a3_10}
        \STATE Apply Algorithm~\ref{alg::mds} on
        $\hat{\mathbf{s}}_{u,v}$, $v=1, \ldots, \alpha$ to solve $\mf S_u$. ($\hat{\mathbf{s}}_{u,v}$ is a subsequence of the $(u,v)$ entry of $\tilde{\mf S}$) \label{a3_11}
    \ENDFOR
    \STATE{Repeat Line~\ref{a3_8} -- \ref{a3_11} on $\hat{\mathbf{q}}_{u,v}$ instead of $\hat{\mathbf{p}}_{u,v}$ to solve $\mf T$.}
\end{algorithmic}
\end{algorithm}

\begin{example}
\label{exam::msr_trans}
Consider the example of $\left[ 6,3,4 \right]$ MSR code in Example~\ref{exam::msr}, where $\alpha=2$. Consider decoding from nodes $1$, $3$ and $4$, i.e., $i_1=4$, $i_2=3$ and $i_3=1$.
The decoder retrieves the sequences $\mathbf{Y}^{4}=\left[\mathbf{y}_{4,1}, \mathbf{y}_{4,2} \right]$, $\mathbf{Y}^{3}=\left[\mathbf{y}_{3,1}, \mathbf{y}_{3,2} \right]$ and $\mathbf{Y}^{1}=\left[\mathbf{y}_{1,1}, \mathbf{y}_{1,2} \right]$, and obtains the shift-XOR equations
\begin{eqnarray*}
\mathbf{y}_{4,1} \!\!&\!=\!&\!\!
{\mathbf{x}_{1}+z^{3}\mathbf{x}_{2}+z^{6}\mathbf{x}_{4}+z^{9}\mathbf{x}_{5}} \\
\mathbf{y}_{4,2} \!\!&\!=\!&\!\!
{\mathbf{x}_{2}+z^{3}\mathbf{x}_{3}+z^{6}\mathbf{x}_{5}+z^{9}\mathbf{x}_{6}} \\
\mathbf{y}_{3,1} \!\!&\!=\!&\!\!
{\mathbf{x}_{1}+z^{2}\mathbf{x}_{2}+z^{4}\mathbf{x}_{4}+z^{6}\mathbf{x}_{5}} \\
\mathbf{y}_{3,2} \!\!&\!=\!&\!\!
{\mathbf{x}_{2}+z^{2}\mathbf{x}_{3}+z^{4}\mathbf{x}_{5}+z^{6}\mathbf{x}_{6}} \\
\mathbf{y}_{1,1} \!\!&\!=\!&\!\!
{\mathbf{x}_{1}+\mathbf{x}_{2}+\mathbf{x}_{4}+\mathbf{x}_{5}} \\
\mathbf{y}_{1,2} \!\!&\!=\!&\!\!
{\mathbf{x}_{2}+\mathbf{x}_{3}+\mathbf{x}_{5}+\mathbf{x}_{6}} .
\end{eqnarray*}
The above system cannot be solved directly using the shift-XOR elimination as it does not satisfy the RID property. We apply Algorithm~\ref{alg::msrdec} to solve the system.

First, the decoder uses the above sequences to calculate  $\hat{\mathbf{c}}_{u,v}$ ($1 \le u\neq v \le 3$), where
\begin{eqnarray*}
\hat{\mathbf{c}}_{1,2}&=&\left(\mathbf{Y}^{4}(\mathbf{\Phi}^{3})^{\top} \right)\left[\range{1}{(L+5)}\right], \\
\hat{\mathbf{c}}_{1,3}&=&\left(\mathbf{Y}^{4}(\mathbf{\Phi}^{1})^{\top} \right)\left[\range{1}{(L+3)}\right],\\
\hat{\mathbf{c}}_{2,1}&=&\mathbf{Y}^{3}(\mathbf{\Phi}^{4})^{\top}[\range{5}{(L+9)}],\\
\hat{\mathbf{c}}_{2,3}&=&\left(\mathbf{Y}^{3}(\mathbf{\Phi}^{1})^{\top} \right)\left[\range{1}{(L+2)}\right],\\
\hat{\mathbf{c}}_{3,1}&=&\mathbf{Y}^{1}(\mathbf{\Phi}^{4})^{\top}[\range{3}{(L+5)}],\\
\hat{\mathbf{c}}_{3,2}&=&\mathbf{Y}^{1}(\mathbf{\Phi}^{3})^{\top}[\range{3}{(L+4)}].
\end{eqnarray*}
By \eqref{equ::generator_msr_1}, we have the system of shift-XOR equations
\begin{equation*}
  \begin{bmatrix}
\mathbf{c}_{2,1} \\
\mathbf{c}_{1,2}
\end{bmatrix} =
\begin{bmatrix}
{1} & {z^{\lambda_{i_2}}} \\
{1} & {z^{\lambda_{i_{1}}}}
\end{bmatrix}
\begin{bmatrix}
\mathbf{p}_{2,1} \\
\mathbf{q}_{2,1}
\end{bmatrix}.
\end{equation*}
Performing the shift-XOR elimination on $\hat{\mf c}_{2,1}$ and $\hat{\mf c}_{1,2}$, we obtain $\mf p_{2,1}$ and $\mf q_{2,1}$. Similarly, we obtain $\mf p_{3,1}$ and $\mf q_{3,1}$ from $\hat{\mf c}_{1,3}$ and $\hat{\mf c}_{3,1}$, and obtain $\mf p_{3,2}$ and $\mf q_{3,2}$ from $\hat{\mf c}_{2,3}$ and $\hat{\mf c}_{3,2}$.
We further generate $\mf p_{v,u}$ and $\mf q_{v,u}$ for $1\leq v <u \leq 3$ by symmetry.

By \eqref{equ::generator_msr_2}, we can form two systems
\begin{equation*}
  \begin{bmatrix}
    \mathbf{p}_{2,1} \\
    \mathbf{p}_{3,1}
  \end{bmatrix}  =
  \begin{bmatrix}
    \mathbf{\Phi}^{i_{2}} \\
    \mathbf{\Phi}^{i_{3}} \\
  \end{bmatrix}
  \begin{bmatrix}
    \tilde{\mf s}_{1,1} \\ \tilde{\mf s}_{1,2}
  \end{bmatrix},
\end{equation*}
and
\begin{equation*}
  \begin{bmatrix}
    \mathbf{p}_{1,2} \\
    \mathbf{p}_{3,2}
  \end{bmatrix}  =
  \begin{bmatrix}
    \mathbf{\Phi}^{i_{1}} \\
    \mathbf{\Phi}^{i_{3}} \\
  \end{bmatrix}
  \begin{bmatrix}
    \tilde{\mf s}_{2,1} \\ \tilde{\mf s}_{2,2}
  \end{bmatrix},
\end{equation*}
solving of which give us $\tilde{\mf s}_{v,u}$, $1\leq v, u\leq 2$.
Then by \eqref{eq:tildes}, we have the system
\begin{equation*}
  \begin{bmatrix}
    \tilde{\mf s}_{1,1} & \tilde{\mf s}_{1,2} \\
    \tilde{\mf s}_{2,1} & \tilde{\mf s}_{2,2}
  \end{bmatrix}
  =
  \begin{bmatrix}
    \mathbf{\Phi}^{i_{1}} \\ \mathbf{\Phi}^{i_{2}}
  \end{bmatrix}
  \begin{bmatrix}
    \mf x_1 & \mf x_2 \\ \mf x_2 & \mf x_3
  \end{bmatrix}.
\end{equation*}
Applying the shift-XOR elimination on $\hat{\mathbf{s}}_{1,1}$ and $\hat{\mathbf{s}}_{2,1}$, we obtain $\mathbf{x}_1$ and $\mathbf{x}_2$. 
Applying the shift-XOR elimination on $\hat{\mathbf{s}}_{1,2}$ and $\hat{\mathbf{s}}_{2,2}$, we obtain $\mathbf{x}_2$ and $\mathbf{x}_3$. 

Executing the same process above using ${\mathbf{q}}_{u,v}$ in place of $\mf p_{u,v}$, we can solve $\mf x_4, \mf x_5, \mf x_6$.
\end{example}

\subsubsection{Complexity Analysis}

\textbf{Time Complexity:} The time complexity of Algorithm~\ref{alg::msrdec} can be calculated based on the time complexity of shift-XOR eliminations performed in the algorithm. First, $k\alpha(\alpha-1)(L+O(nd))=k(k-1)(k-2)(L+O(nd))$ XOR operations are needed for computing $\hat{\mf c}_{u,v}$ and $\hat{\mathbf{c}}_{v,u}$ for $1\leq u<v\leq k$. Solving $\mathbf{p}_{u,v}$ and $\mathbf{q}_{u,v}$ for $u=2,\ldots,k$ and $v=1,\ldots,u-1$ costs $\alpha (\alpha +1)(L+O(nd)) = k(k-1)(L+O(nd))$ XOR operations.
 Then solving $\mathbf{S}$ and $\mathbf{T}$ costs $4\alpha ^{2}(\alpha-1)(L+O(nd))=4(k-1)^{2}(k-2)(L+O(nd))$ XOR operations.
Totally, the time complexity $T$ for decoding of MSR codes is
\begin{eqnarray*}
T &=& k\left(k-1\right)\left(k-2\right)(L+O(nd))+k\left(k-1\right)(L+O(nd))+4(k-1)^{2}(k-2)(L+O(nd)) \\
&=& \left(k-1 \right)^2 \left(5k-8\right)L+O(nk^3d).
\end{eqnarray*}

\textbf{Space Complexity:}
In Algorithm~\ref{alg::msrdec}, the $2\alpha^2$ sequences $\mf p_{u,v}$, $\mf p_{v,u}$ take  $2\alpha ^2(L+O(nd))$ bits storage, which is the largest space cost of the algorithm during an execution.
The message sequences has $\alpha (\alpha+1) L$ bits, so the auxiliary space is $\alpha(\alpha-1)L+O(nd^3)$ bits.

\textbf{Bandwidth:}
As the number of bits transmitted to the decoder from node $i_{j}$ is $\alpha (L+t_{i_{j},\alpha}+\lambda_{i_{j}})$, the total number of transmitted bits is $k\alpha L+\alpha \sum_{j=1}^{k}(t_{i_{j},\alpha}+\lambda_{i_{j}})$.
There are $k\alpha L$ bits in the message sequences, and hence the bandwidth overhead is $\alpha \sum_{j=1}^{k}(t_{i_{j},\alpha}+\lambda_{i_{j}})=O(nk^2d)$.

\subsection{Repair Scheme of Shift-XOR MSR Codes}
\label{sec::msr_repair}

This section introduces our repair scheme for an $[n,k,d]$ shift-XOR MBR code, where $d=2k-2$ and $\alpha=k-1$.
Define an $\alpha\times d$ matrix
$\mf X = (\mf x_{i,j})$ with $\mf x_{i,j}=\mathbf{\Phi}^{i}\mathbf{S}_j$ and $\mf x_{i,j+\alpha}=\mathbf{\Phi}^{i}\mathbf{T}_j$ for $1\leq i,j\leq \alpha$.
Suppose that node $i$ fails. Our repair scheme generates a new storage
node that stores the same $\alpha$ sequences at node $i$: 
\begin{equation}
    \label{eq:msr_repair}
    \mathbf{Y}^i = \mathbf{\Psi}^{i}\mathbf{M} = \mathbf{\Phi}^i\mathbf{S}+z^{\lambda_i}\mathbf{\Phi}^i\mathbf{T}=
    \mf X^i_{\range{1}{\alpha}}+ z^{\lambda_i} \mf X_{\range{(\alpha+1)}{2\alpha}}^i.
\end{equation}
Recall that each storage node $j\in 1:n$ stores the sequences $\mathbf{\Psi}^{j}\mf M$, and hence can compute locally the sequence
\begin{equation}\label{eq:rmsr}
  \mathbf{r}_j=\mathbf{\Psi}^{j}\mathbf{M}\left(\mathbf{\Phi}^i\right)^{\top}=\mathbf{\Psi}^{j} (\mf X^i)^\top,
  % \begin{bmatrix}
  %   \mf X_S^i & \mf X_T^i
  % \end{bmatrix}^\top,
\end{equation}
which is a shift-XOR equation of $\mf X^i$.
By $\mf r_j$ from any $d$ nodes $j\neq i$, we can solve $\mf X^i$ using the shift-XOR elimination, and then calculate $\mf Y^i$ by \eqref{eq:msr_repair}.

Specifically, the repair scheme includes two stages: the transmission stage and the decoding stage.
In the transmission stage, $d$ {helper nodes} are chosen to repair node $i$, which have the indices 
 $i_1, i_2, \ldots, i_d$ where $i_{1}>i_{2}>\cdots >i_{d}$.
Each helper node $i_{v}$  ($1 \le v \le d$) transmits 
\begin{equation}
\label{eq:mhat_msr_repair}
\hat{\mathbf{r}}_v =
\mathbf{r}_{i_v}[t_{i_v,v}+(\range{1}{(L+t_{i,\alpha})})]
\end{equation}
to the new node $i$ for repairing, where the sequence  $\mathbf{r}_{i_v}$ is defined in \eqref{eq:rmsr}.
%Note that a sequence in $\mathbf{Y}^i$ has $L+t_{i,\alpha}+\lambda_i$ bits, and hence $\hat{\mathbf{r}}_v$ has exactly the same number of bits as a sequence to repair.
In the decoding stage, the new node $i$ performs the shift-XOR elimination on $\hat{\mathbf{r}}_v$, $v=1,\ldots,d$ to decode $\mf X^i$, and then calculate $\mathbf{Y}^i$ by~\eqref{eq:msr_repair}.
% A pseudocode of the decoding stage is shown in Algorithm~\ref{alg::msrRepair}.

% \begin{algorithm}[t]
% \caption{Repair Algorithm for Shift-XOR MSR Codes}
% \label{alg::msrRepair}
% \begin{algorithmic}[1]
% \REQUIRE sequences ${{\hat{\mathbf{r}}}_{j}}$  ($1\leq j \leq d$).

% \ENSURE $\mathbf{Y}^{i}$. %, stored in ${{\hat{\mathbf{m}}}_{j}}$  ($1\leq j \leq d$).

% %\textbf{Step~1:} \textit{(Calculate $\mathbf{M}(\mathbf{\Phi}^{i})^{\top}$)}

% \STATE {Apply Algorithm~\ref{alg::mds} on ${{\hat{\mathbf{r}}}_{v}}$,  $v = 1,\ldots, d$ to decode $\mf X^i$.}
% %\textbf{Step~2:} \textit{(Calculate $\mathbf{Y}^{i}$)}

% \FOR {$j\leftarrow 1:\alpha$}
%         \STATE {$\mathbf{x}_{i,j} \oplus  \leftarrow z^{\lambda_i} \mathbf{x}_{i,j+\alpha}$}
% \ENDFOR
% \end{algorithmic}
% \end{algorithm}

\begin{example}
Consider the $\left[6,3,4 \right]$ MSR Code studied in Example~\ref{exam::msr}.
This example will show the transmission and decoding stages to repair node $3$ from helper nodes $1$, $2$, $4$ and $5$, i.e., $i_{1}=5$, $i_{2}=4$, $i_{3}=2$, $i_{4}=1$ and $i = 3$. 
The sequences transmitted to node $i$ from the $4$ helper nodes are
\begin{eqnarray}
\hat{\mathbf{r}}_{1} &=&\mathbf{r}_{5}\left[\range{1}{(L+2)} \right] = (\mathbf{Y}^{5}(\mathbf{\Phi}^{3})^{\top})\left[\range{1}{(L+2)} \right], \label{eq:r5} \\
\hat{\mathbf{r}}_{2} &=&\mathbf{r}_{4}\left[\range{4}{(L+5)} \right] = (\mathbf{Y}^{4}(\mathbf{\Phi}^{3})^{\top})\left[\range{4}{(L+5)} \right], \label{eq:r4} \\
\hat{\mathbf{r}}_{3} &=&\mathbf{r}_{2}\left[\range{3}{(L+4)} \right] = (\mathbf{Y}^{2}(\mathbf{\Phi}^{3})^{\top})\left[\range{3}{(L+4)} \right], \label{eq:r2} \\
\hat{\mathbf{r}}_{4} &=&\mathbf{r}_{1} = \mathbf{Y}^{1}(\mathbf{\Phi}^{3})^{\top}. \label{eq:r1}
\end{eqnarray}

By~\eqref{eq:rmsr}, we have the system
\begin{equation*}
    \begin{bmatrix}
    \mathbf{r}_5\\
    \mathbf{r}_4\\
    \mathbf{r}_2\\
    \mathbf{r}_1\\
    \end{bmatrix}=\begin{bmatrix}
    \mathbf{\Phi}^5\\
    \mathbf{\Phi}^4\\
    \mathbf{\Phi}^2\\
    \mathbf{\Phi}^1\\
    \end{bmatrix}\left(\mathbf{X}^3\right)^{\top}.
\end{equation*}
%where $\mathbf{X}^3=\mathbf{\Phi}^3\left[\mathbf{S}~\mathbf{T}\right]$.
Applying the shift-XOR elimination on $\hat{\mathbf{r}}_1$, $\hat{\mathbf{r}}_2$, $\hat{\mathbf{r}}_3$ and $\hat{\mathbf{r}}_4$, we can obtain ${\mathbf{X}}^3$, and hence solve $\mathbf{Y}^3$ by~\eqref{eq:msr_repair}.
\end{example}

\textbf{Time Complexity:}
At each helper node $i_v$ ($v=1,\ldots,d$), computing $\hat{\mathbf{r}}_{i_v}$ costs  $(\alpha -1)(L+O(nd))$ XOR operations.
So there are totally $d\left(\alpha-1\right)(L+O(nd))=d(\frac{d}{2}-1)(L+O(nd))$ XOR operations at all the helper nodes.
At the new node $i$, solving $\mathbf{X}^i$ costs $d(d-1)(L+O(nd))$ XOR operations by the shift-XOR elimination. Computing $\mathbf{Y}^i$ costs $\alpha (L+O(n\alpha))=\frac{d}{2}(L+O(n\alpha))$ XOR operations. So the total time complexity at the new node $i$ is $d(d-1/2)(L+O(nd))$.
The overall time complexity among all the involved nodes is $\frac{3}{2}(d-1)dL+O(nd^3)$.

\textbf{Bandwidth:} As the number of bits transmitted from node $i_v$ is $L+t_{i,\alpha}$, so the total number of bits transmitted is $d(L+t_{i,\alpha})=dL+O(nd^2)$.

\textbf{Space Complexity:}
The storage space of $d(L+t_{i,\alpha})=d(L+O(nd))$ bits is required to store the bits retrieved from the helper nodes. As the shift-XOR elimination can be implemented in-place, no extra storage space is required to store the intermediate results $\mathbf{X}^i$. 
The total length of the repaired sequences is $\alpha (L+\lambda_{i}+t_{i,\alpha})=\frac{d}{2}(L+O(nd))$. So the auxiliary space required for intermediate XOR results is $\frac{d}{2} (L+O(nd))$.

\section{Extensions to other PM-Constructed Codes}
\label{sec::application}
The decompositions of our decoding and repair schemes discussed in the
previous two sections depend mostly on the PM construction, and have
little correlation to the shift and XOR operations. Therefore, similar
decomposition may be possible for the decoding and repair of other
regenerating codes based on the PM construction. In this section, we study the extensions of our decoding and repair schemes to the finite-field PM
codes~\cite{rashmi2011optimal,LinCHA15} and the cyclic-shift PM
codes~\cite{HouSCL16}.

\subsection{Extension to Finite-Field PM Codes}
The finite-field PM codes in~\cite{rashmi2011optimal} use finite field operations. Suppose the entries of a sequence are elements from a finite field $\mathbb{F}$. Same as the setting in Section~\ref{sec:gen_enc}, using finite field operations, we define
\begin{equation*}
  \mathbf{y}_{i,j} = \sum_{u=1}^d \psi_{i,u} \mathbf{m}_{u,j}, \quad 1 \le i \le n, 1 \le j \le \alpha,
\end{equation*}
where $\psi_{i,u}\in \mathbb{F}$, and $\mf y_{i,j}$ and $\mf m_{u,j}$ are sequences of $L$ bits, or $L$ symbols from $\mathbb{F}$. Denoting $\mathbf{\Psi} = (\psi_{i,j})$, called the generator matrix, the encoding follows the same form of~\eqref{equ::form}:
\begin{equation}
  \label{eq:1}
  \mathbf{Y} = \mathbf{\Psi} \mathbf{M}.
\end{equation}
As Gaussian elimination can solve systems of linear equations over finite fields, we can use Gaussian elimination in place of shift-XOR elimination to build decode/repair schemes for the finite-field PM codes.

\subsubsection{Decoding Scheme of Finite-Field MBR Codes}
\label{sec::app_ff}
For the finite-field MBR codes in~\cite{rashmi2011optimal}, $\mathbf{\Psi}$ in \eqref{eq:1} is of the form
\begin{equation*}
  \mathbf{\Psi}=\left[\mathbf{\Phi} \quad \mathbf{\Delta}\right],
\end{equation*}
where $\mathbf{\Phi}=\left(\phi_{i,j}\right)$ and $\mathbf{\Delta}$ are $n\times k$ and $n\times (d-k)$ matrices respectively and satisfy: 1) any $d$ rows of $\mathbf{\Psi}$ are linearly independent; 2) any $k$ rows of $\mathbf{\Phi}$ are linearly independent. Specifically, the Vandermonde matrix and Cauchy matrix satisfy the above two conditions~\cite{rashmi2011optimal}.
The message matrix is of the form
\begin{equation*}
 \mathbf{M}=
 \begin{bmatrix}
 \mathbf{S} & \mathbf{T}\\
 \mathbf{T}^\top & \mathbf{O}
\end{bmatrix},
\end{equation*}
where $\mf S$ is a $k\times k$ symmetric matrix and $\mf T$ is a $k\times (d-k)$ matrix. There are totally $\frac{1}{2}(k+1)k+k(d-k)$ message sequences.
In the decoding algorithm of~\cite{rashmi2011optimal}, the $d k$ sequences stored at $k$ nodes are retrieved for decoding, so that the decoding bandwidth overhead is $\frac{1}{2}k(k-1)$ sequences. 

Due to the same matrix form as the shift-XOR MBR codes, we may wonder whether it is possible to derive a similar decoding algorithm as in Section~\ref{sec::mbr_decode} with zero bandwidth overhead. We show it is possible when $\mf \Phi$ satisfies the further requirement that for any $k$ rows of $\mathbf{\Phi}$, all the leading principal submatrices are full rank. It is noted that the Vandermonde matrix and Cauchy matrix used in~\cite{rashmi2011optimal} also satisfy the requirement.

To illustrate the algorithm, suppose the first $k$ storage nodes are used for decoding. The decoding using other choice of nodes is similar. The decoder  first retrieves 
\begin{equation*}
\mathbf{Y}_{(k+1):d}^{1:k}=\mathbf{\Phi}^{1:k} \mathbf{T}.
\end{equation*}
As any $k$ rows of $\mathbf{\Phi}$ are linearly independent, $\mathbf{T}$ can be decoded using Gaussian elimination. 
After decoding $\mathbf{T}$, we continue to decode the $k$ columns of $\mathbf{S}$ sequentially with the column indices in descending order.
For $u = k, k-1, \ldots, 1$,
\begin{IEEEeqnarray*}{rCl}
  \mathbf{Y}_u^{1:u}
  & = &
  \mathbf{\Phi}^{1:u}_{1:u} \mathbf{S}^{\range{1}u}_u + \mathbf{\Phi}^{1:u}_{\range{(u+1)}{k}} \mathbf{S}^{\range{(u+1)}{k}}_u +
  \mathbf{\Delta}^{1:u} (\mathbf{T}^u)^\top,
\end{IEEEeqnarray*}
i.e.,
\begin{equation*}
  \mathbf{Y}_u^{1:u} - \mathbf{\Phi}^{1:u}_{\range{(u+1)}{k}} \mathbf{S}^{\range{(u+1)}{k}}_u
  - \mathbf{\Delta}^{1:u} (\mathbf{T}^u)^\top = \mathbf{\Phi}^{1:u}_{1:u} \mathbf{S}^{\range{1}u}_u.
\end{equation*}
After substituting $\mathbf{T}$ and $\mathbf{S}^{\range{(u+1)}{k}}_u = \mathbf{S}^u_{\range{(u+1)}{k}}$, the LHS of the above equation is known and $\mathbf{S}^{\range{1}u}_u$ can be decoded by Gaussian elimination as $\mathbf{\Phi}^{1:u}_{1:u}$ has rank $u$.
Hence, $\mathbf{S}$ can be decoded column by column. % by executing Gaussian elimination.

The above decoding scheme has the same asymptotic time/space/bandwidth complexity as the scheme for the shift-XOR MBR codes, and the decoder retrieves exactly the same number of bits as the message sequences.

\subsubsection{Decoding Scheme of Finite-Field MSR Codes}
\label{sec:decode-ff-msr}
For finite-field MSR codes with $d=2k-2$ and $\alpha=k-1$ in~\cite{rashmi2011optimal}, the generator matrix $\mathbf{\Psi}$ is of the form
\begin{equation*}
  \mathbf{\Psi}=\left[\mathbf{\Phi}\quad \mathbf{\Lambda}\mathbf{\Phi}\right],
\end{equation*}
where $\mathbf{\Phi}$ is an $n\times \alpha$ matrix and $\mathbf{\Lambda}=\text{diag}(\lambda_1,\ldots,\lambda_n)$ is an $n\times n$ diagonal matrix and satisfy: 1) any $d$ rows of $\mathbf{\Psi}$ are linearly independent; 2) any $\alpha$ rows of $\mathbf{\Phi}$ are linearly independent; (3) the $n$ diagonal elements in $\mathbf{\Lambda}$ are distinct.
The message matrix is of the form 
\begin{equation*}
    \mathbf{M}=
    \begin{bmatrix}
    \mathbf{S}\\
    \mathbf{T}
    \end{bmatrix}.
\end{equation*}
The coded sequences stored are obtained by 
\begin{equation*}
    \mathbf{Y}=\mathbf{\Psi}\mathbf{M}.
\end{equation*}
The $i$-th row of $\mathbf{Y}$ are stored at node $i$, i.e., $\mathbf{Y}^i=\mathbf{\Phi}^{i}\mathbf{S}+\lambda_i\mathbf{\Phi}^i\mathbf{T}$. 

Assume that the decoder has access to $k$ nodes $i_u$ for $1
\leq u\leq k$.
For decoding, node $i_u$ transmits $\mathbf{Y}^{i_u}$ to the decoder. As discussed in Section~\ref{sec::msr_decode}, denote for $1\leq u \neq v \leq k$,
\begin{IEEEeqnarray*}{rCl}
  \mf c_{u,v} & = & \mf Y^{i_u} (\mathbf{\Phi}^{i_{v}})^\top, \\
  \mf p_{u,v} & = & \mf\Phi^{i_u} \mf S (\mathbf{\Phi}^{i_{v}})^\top, \\
  \mf q_{u,v} & = & \mf\Phi^{i_u} \mf T (\mathbf{\Phi}^{i_{v}})^\top.
\end{IEEEeqnarray*}

$\mathbf{S}$ and $\mathbf{T}$ can be solved by the following two steps.

\emph{Step 1:}
For $1\leq v<u\leq k$, we have the linear systems
\begin{equation*}
\begin{bmatrix}
\mathbf{c}_{u,v} \\
\mathbf{c}_{v,u}
\end{bmatrix} =\begin{bmatrix}
\mathbf{p}_{u,v}+{\lambda_{i_u}}\mathbf{q}_{u,v}\\
\mathbf{p}_{v,u}+{\lambda_{i_v}}\mathbf{q}_{v,u}\\
\end{bmatrix}=
\begin{bmatrix}
\mathbf{p}_{u,v}+{\lambda_{i_u}}\mathbf{q}_{u,v}\\
\mathbf{p}_{u,v}+{\lambda_{i_v}}\mathbf{q}_{u,v}\\
\end{bmatrix}=
\begin{bmatrix}
{1} & {\lambda_{i_u}} \\
{1} & {\lambda_{i_v}}
\end{bmatrix}
\begin{bmatrix}
\mathbf{p}_{u,v} \\
\mathbf{q}_{u,v}
\end{bmatrix}.
\end{equation*}
Gaussian elimination is performed on $\mf c_{u,v}$ and $\mf c_{v,u}$ to solve $\mf p_{u,v}$ and $\mf q_{u,v}$.

\emph{Step 2:}
Define an $\alpha\times \alpha$ matrix $\tilde{\mathbf{S}} = (\tilde{\mf s}_{v,u})$ as 
\begin{equation}
\label{eq:stilde_fmsr}
  \tilde{\mathbf{S}} = %\mathbf{\Phi}^{i_{v}}
  \begin{bmatrix}
    \mathbf{\Phi}^{i_{1}} \\ \vdots \\ \mathbf{\Phi}^{i_{\alpha}}
  \end{bmatrix}
  \mf S.
\end{equation}
Due to the symmetry of $\mf S$, $(\tilde{\mathbf{S}}^v)^\top = \mf S (\mf\Phi^{i_v})^\top$.
%Recall that $k=\alpha+1$.
For each $v\in 1:\alpha$, form the $\alpha\times \alpha$ linear system
\begin{equation*}
\begin{bmatrix}
\mathbf{p}_{1,v} \\
\vdots \\
\mathbf{p}_{v-1,v} \\
\mathbf{p}_{v+1,v} \\
\vdots \\
\mathbf{p}_{k,v}
\end{bmatrix}  =
\begin{bmatrix}
\mathbf{\Phi}^{i_{1}} \\
\vdots \\
\mathbf{\Phi}^{i_{v-1}} \\
\mathbf{\Phi}^{i_{v+1}} \\
\vdots \\
\mathbf{\Phi}^{i_{k}} \\
\end{bmatrix}
(\tilde{\mathbf{S}}^{v})^{\top},
\end{equation*}
so that $(\tilde{\mathbf{S}}^{v})^{\top}$ can be solved by performing Gaussian  elimination. 
After solving $\tilde{\mathbf{S}}$, we further solve the linear system~\eqref{eq:stilde_fmsr}.
Hence, $\mf S_u$ can be decoded by Gaussian elimination on $\tilde{\mathbf{s}}_{v,u}, v=1,\ldots, \alpha$.

$\mf T$ can be solved by the same procedure and hence is omitted. 

For our method, storage space of $2\alpha\times \alpha L=2(k-1)^2L$ symbols are needed;  while in~\cite{rashmi2011optimal} the decoding needs space of $2k(k-1)L$ symbols. Hence, we can reduce space of $2(k-1)L$ symbols. In addition, the computation of $\mathbf{Y}^{i_u}(\mathbf{\Phi}^{i_u})^{\top}$ are omitted in our method, such that $\alpha$ multiplications of a vector and a sequence can be reduced.

In~\cite{LinCHA15}, finite-field MSR codes for $d\geq 2k-2$ are constructed, where the decoding procedure involves the decoding of the finite-field PM MSR codes with $d=2k-2$. Therefore, our decoding scheme mentioned above can also be substituted into the decoding of these MSR codes for $d\geq 2k-2$, and reduce the corresponding decoding complexity.

\subsection{Extension to Cyclic-shift Regenerating Codes}

The cyclic-shift regenerating codes in~\cite{HouSCL16} employ a cyclic-shift operation defined as
\begin{equation*}
\left(z_{c}^t \mathbf{a} \right) \left[ l \right] =
\begin{cases}
\mathbf{a}  \left[ l+L-t \right], & 1\le l \le t,\\
 \mathbf{a}  \left[ l-t \right] , & t < l \le L.
\end{cases}
\end{equation*}
Same as the setting in Section~\ref{sec:gen_enc}, let
\[
{{\mathbf{y}}_{i,j}}=\sum\limits_{u=1}^{d}{{{z}_{c}^{{{t}_{i,u}}}}{{\mathbf{m}}_{u,j}}}, \quad 1 \le i \le n, 1 \le j \le \alpha,
\]
where $t_{i,u}\geq 0$ are integers.
Denoting $\mathbf{\Psi} = (z_{c}^{t_{i,j}})$, the encoding follows the same form as~\eqref{equ::form}.

A system of cyclic-shift equations can be expressed as
\begin{equation*}
\begin{bmatrix}
{\mathbf{y}_{1}} \\
{\mathbf{y}_{2}}\\
{\vdots}\\
{\mathbf{y}_{k}} \\
\end{bmatrix}
=\mathbf{\Psi }
\begin{bmatrix}
{\mathbf{x}_{1}} \\
{\mathbf{x}_{2}} \\
{\vdots} \\
 {\mathbf{x}_{k}} \\
\end{bmatrix},
\end{equation*}
where $\det(\mathbf{\Psi})$ has an inverse element in $\mathbb{F}_{2}[z]/(1+z+\cdots +z^{L-1})$. When $\mathbf{\Psi}$ is a Vandermonde matrix with $k-1$ strictly less than all divisors of $L$ which are not equal to $1$, the system can be solved using the LU method~\cite{HouSCL16}.

Similar to shift-XOR codes in Section~\ref{sec::mbr} and Section~\ref{sec::msr}, the decoding and repair of the cyclic-shift codes can be decomposed into a sequence of systems of cyclic-shift equations.
When $n-1$ (where $n$ is the number of storage nodes) is strictly less than all divisors of $L$ which are not equal to $1$, the sequence of systems can be solved by the LU method. The decoding and repair schemes built in this way have the same asymptotic complexity as that of our shift-XOR codes.

\section{Concluding Remarks}
\label{sec::conclude}

One technical contribution of this paper is an efficient algorithm called shift-XOR elimination to solve a system of shift-XOR equations satisfying the RID property. Our algorithm consumes the exactly same number of XOR operations for decoding as encoding the input subsequences, and can be implemented in-place with only a small constant number of auxiliary integer variables. The shift-XOR elimination has the potential to be applied to and simplified the decoding costs of a range of codes based on shift-XOR operations. 

For shift-XOR regenerating codes, the decoding/repair schemes are decomposed into a sequence of systems of shift-XOR equations. Our decoding/require schemes have much lower computation costs than the existing schemes for the shift-XOR regenerating codes, and demonstrate better or similar computation costs compared with the regenerating codes based on cyclic-shift and XOR operations. Our results provide a further evidence that shift and XOR operations can help to design codes with low computation costs. 

Though we only studied the bit-wise shifts in this paper, our
algorithms can be extended to byte-wise or word-wise shifts to utilize
multi-bit computation devices in parallel.

We are motivated to further explore the potential of shift-XOR codes. In one direction, we may extend the code constructions based on finite-field/cyclic-shift operations (e.g., \cite{LinCHA15,KuriharaK13,ElyasiM19,Elyasi19Cascade,YeB17a,YeB17,LiTT18}) to ones using shift and XOR. In another direction, we may investigate non-RID generator matrices, which may have lower storage overhead.

\appendix[Proof of Theorem~\ref{theo::mds}]
\label{sec::proof_MDS}

Here we prove Theorem~\ref{theo::mds}, which concerns
a $k\times k$ system of shift-XOR equations~\eqref{equ::mds_enc}, where $\mathbf{\Psi} = (z^{t_{i,j}})$ satisfies the RID property in Definition~\ref{def:rip}. Recall $L_b$ defined in \eqref{eq:tb}.

\begin{lemma}\label{lemma:tbb}
For integers $1\leq u < v < k$,
\begin{equation*}
 t_{k-v,v+1} - t_{k-v,u}  < \sum_{b=u}^v L_b < t_{k-u,v+1} - t_{k-u,u}.
\end{equation*}
\end{lemma}
\begin{IEEEproof}
  The lemma can be proved by applying the RID property. On the one hand,
  \begin{IEEEeqnarray*}{rCl}
    \sum_{b=u}^v L_b & = & \sum_{b=u}^v (t_{k-b,b+1}-t_{k-b,b}) \\
    & < & \sum_{b=u}^v (t_{k-u,b+1}-t_{k-u,b}) \\
    & = & t_{k-u,v+1} - t_{k-u,u}.
  \end{IEEEeqnarray*}
  On the other hand,
  \begin{IEEEeqnarray*}{rCl}
    \sum_{b=u}^v L_b & = & \sum_{b=u}^v (t_{k-b,b+1}-t_{k-b,b}) \\
    & > & \sum_{b=u}^v (t_{k-v,b+1}-t_{k-v,b}) \\
    & = & t_{k-v,v+1} - t_{k-v,u}.
  \end{IEEEeqnarray*}
\end{IEEEproof}

Now we start to prove Theorem~\ref{theo::mds}.
We inductively show that all the bits to solve in each iteration depend on only the previous solved bits. We use $l_i$ to denote the number of bits solved in $\bx_i$, which are zero initially. For $k=1$, the shift-XOR elimination is successful without using back substitution. We consider $k>1$ in the following proof.

Firstly, for an iteration $\iters$ in $1:L_1$, we see that
\begin{equation*}
%  \label{eq:2}
  \bx_1[\iters]  = \hat\bx_1[\iters] + \sum_{u=2}^{k}\bx_{u}[\iters+t_{k,1}-t_{k,u}]
\end{equation*}
As $\iters\leq L_1 = t_{k-1,2}-t_{k-1,1}$, we have $\iters+t_{k,1} - t_{k,u} < t_{k,2}-t_{k,1} + t_{k,1} - t_{k,u} = t_{k,2} - t_{k,u} \leq 0$ for $u\geq 2$ due to the RID property. Hence $\bx_1[\iters]  = \hat\bx_1[\iters]$ so that $\bx_1[\iters]$ can be solved. After iteration $L_1$, we have $l_1=L_1$ and $l_i=0$ for $i>1$. 

For certain $2\leq b \leq k$, fix an iteration $\iters$ in $\sum_{b'=1}^{b-1}L_{b'}+(1:L_b)$ and an index $i$ in $1:b$. We assume that the algorithm runs successfully to iteration $\iters$ with $\bx_{u}[\iters-\sum_{b'=1}^{u-1}L_{b'}]$, for all $u<i$ solved, i.e.,
\begin{equation} \label{eq:lu}
  l_u =
  \begin{cases}
    \iters - \sum_{b'=1}^{u-1}L_{b'}, & 1 \leq u < i, \\
    \iters-1 - \sum_{b'=1}^{u-1}L_{b'}, & i \leq u \leq b, \\
    0 & u>b.
  \end{cases}
\end{equation}
To check whether $\bx_i[l_i+1]$ can be solved or not, we write by \eqref{eq:xhat2}
\begin{equation}
  \label{eq:2x}
  \bx_i[l_i+1]  = \hat\bx_i[l_i+1] + \sum_{u\neq i}\bx_{u}[l_i+1+t_{k-i+1,i}-t_{k-i+1,u}].
\end{equation}
% Using the RID property and Lemma~\ref{lemma:tbb}, we have
We can check the second term on the RHS is solved as follows:
\begin{enumerate}
\item For $1\leq u \leq i-1$, $\bx_{u}[l_i+1+t_{k-i+1,i}-t_{k-i+1,u}]$ has been solved as 
\begin{IEEEeqnarray*}{rCl}
  l_i+1 + t_{k-i+1,i}-t_{k-i+1,u}
  & = & s - 1 - \sum_{b'=1}^{i-1}L_{b'} + 1 + t_{k-i+1,i}-t_{k-i+1,u} \\
  & = & l_u - \sum_{b'=u}^{i-1}L_{b'} + t_{k-i+1,i}-t_{k-i+1,u} \\
  & \leq & l_u,
\end{IEEEeqnarray*}
where the first two equalities are obtained by substituting the formula in \eqref{eq:lu}, and the inequality is obtained by Lemma~\ref{lemma:tbb}.
\item For $i+1\leq u \leq b$, $\bx_{u}[l_i+1+t_{k-i+1,i}-t_{k-i+1,u}]$ has been solved as 
\begin{IEEEeqnarray*}{rCl}
  l_i+1 + t_{k-i+1,i}-t_{k-i+1,u}
  & = & s - 1 - \sum_{b'=1}^{i-1}L_{b'}+1 + t_{k-i+1,i}-t_{k-i+1,u} \\
  & = & l_u + 1 + \sum_{b'=i}^{u-1}L_{b'} + t_{k-i+1,i}-t_{k-i+1,u} \\
  & \leq & l_u + t_{k-i,u} - t_{k-i,i}  + t_{k-i+1,i}-t_{k-i+1,u} \\
  & < & l_u,
\end{IEEEeqnarray*}
where the first two equalities are obtained by substituting the formula in \eqref{eq:lu},  the first inequality is obtained by Lemma~\ref{lemma:tbb}, and the last inequality follows from the RID property.
\item For $b<u\leq k$, $\bx_{u}[l_i+1+t_{k-i+1,i}-t_{k-i+1,u}] = 0$ as
\begin{IEEEeqnarray*}{rCl}
  l_i+1 + t_{k-i+1,i}-t_{k-i+1,u} & = & \iters - \sum_{b'=1}^{i-1}L_{b'} + t_{k-i+1,i}-t_{k-i+1,u} \\
  & \leq & \sum_{b'=1}^{u-1}L_{b'} - \sum_{b'=1}^{i-1}L_{b'} + t_{k-i+1,i}-t_{k-i+1,u} \\
  & = & \sum_{i'=i}^{u-1}L_{i'} + t_{k-i+1,i}-t_{k-i+1,u} \\
  & < & 0,
\end{IEEEeqnarray*}
where the first equality follows from \eqref{eq:lu}, the first inequality follows from $s\leq \sum_{b'=1}^{b}L_{b'}\leq \sum_{b'=1}^{u-1}L_{b'}$, and the second inequality is obtained by Lemma~\ref{lemma:tbb} and the RID property.
\end{enumerate}
Therefore, all terms on the RHS of \eqref{eq:2x} are known and hence $\bx_i[l_i+1]$ can be solved. 
The proof of the theorem is completed.

% Last, fix an iteration $l$ in $\sum_{j'=1}^{k-1}L_{j'}+(1:L)$ and an index $i$ in $1:k$. We assume that the algorithm runs successfully to iteration $l$ with $\bx_{u}[l-\sum_{b'}^{u-1}L_{b'}]$, for all $u<i$ solved, i.e.,
% \begin{equation*}
%   l_u =
%   \begin{cases}
%     l - \sum_{b'=1}^{u-1}L_{b'} & 1 \leq u < i, \\
%     l-1 - \sum_{b'=1}^{u-1}L_{b'} & i \leq u \leq k.
%   \end{cases}
% \end{equation*}
% Now we check whether $\bx_i[l_i+1]$ as expressed in \eqref{eq:2x} can be solved or not. 
% Using the RID property and Lemma~\ref{lemma:tbb}, we can derive
% \begin{enumerate}
% \item for $1\leq u \leq i-1$, $\bx_{u}[l_i+1+t_{k-i+1,i}-t_{k-i+1,u}]$ has been solved as 
% \begin{IEEEeqnarray*}{rCl}
%   l_i+1 + t_{k-i+1,i}-t_{k-i+1,u} & = & l_u - \sum_{b'=u}^{i-1}L_{b'} + t_{k-i+1,i}-t_{k-i+1,u} \\
%   & \leq & l_u;
% \end{IEEEeqnarray*}
% \item for $i+1\leq u \leq k$, $\bx_{u}[l_i+1+t_{k-i+1,i}-t_{k-i+1,u}]$ has been solved as 
% \begin{IEEEeqnarray*}{rCl}
%   l_i+1 + t_{k-i+1,i}-t_{k-i+1,u} & = & l_u + 1 + \sum_{b'=i}^{u-1}L_{b'} + t_{k-i+1,i}-t_{k-i+1,u} \\
%   & < & l_u;
% \end{IEEEeqnarray*}
% \end{enumerate}
% Therefore, all terms on the RHS of \eqref{eq:2x} are known and hence $\bx_i[l_i+1]$ can be solved.

\ifCLASSOPTIONcaptionsoff
  \newpage
\fi

%    %\end{thebibliography}
    \bibliographystyle{IEEEtran}
%    % argument is your BibTeX string definitions and bibliography database(s)
    \bibliography{IEEEabrv,regenerate}
%    % biography section

%

\end{document}